\documentclass{nature}

\usepackage{aas_macros}
\usepackage{amssymb}
\usepackage{graphicx}
\usepackage{color}
\usepackage[normalem]{ulem} 
\usepackage{soul}
\usepackage{url}
\usepackage[colorlinks]{hyperref}

\usepackage{subfig}
\usepackage[font=footnotesize]{caption}

\usepackage{floatrow}
\DeclareFloatFont{tiny}{\scriptsize}
\title{Massive Stars as Major Factories of Galactic Cosmic Rays}

\newcounter{firstbib}

\begin{document}

\maketitle

\author{
Felix~Aharonian$^{1,2,3}$,
Ruizhi Yang$^{2}$,
Emma de O\~na Wilhelmi$^{4,5,6}$
}

\begin{affiliations}
\small
\item Dublin Institute for Advanced Studies, 31 Fitzwilliam Place, Dublin 2, Ireland 
\item Max-Planck-Institut f\"ur Kernphysik, P.O. Box 103980, D 69029 Heidelberg, Germany 
\item Gran Sasso Science Institute, 7 viale Francesco Crispi, 67100 L'Aquila,  Italy
\item Institute of Space Sciences (ICE/CSIC), Campus UAB, Carrer de Can Magrans s/n, 08193 Barcelona, Spain
\item Institut d' Estudis Espacials de Catalunya (IEEC), 08034 Barcelona, Spain
\item Deutsches Elektronen Synchrotron DESY, 15738 Zeuthen, Germany
\end{affiliations}

\hfill

\begin{abstract}
The identification of major contributors to the locally observed fluxes of  Cosmic Rays (CRs) is a prime objective towards the resolution of the long-standing enigma of CRs.
We report on  a compelling similarity of the energy and radial distributions of multi-TeV CRs extracted from observations of very high energy (VHE) $\gamma$-rays towards the Galactic Center (GC) and two prominent clusters of young massive stars, Cyg~OB2 and Westerlund~1.  This resemblance we interpret as a hint that CRs responsible for the diffuse VHE  $\gamma$-ray emission from the GC are accelerated by the ultracompact stellar clusters located in the heart of GC.  
The derived $1/r$  decrement of the CR density with the distance from a  star cluster is a distinct signature of continuous,  over a few million years, CR injection into the interstellar medium.  The lack of brightening of the  $\gamma$-ray images toward the stellar clusters excludes the leptonic origin of $\gamma$-radiation. 
The hard,   $\propto E^{-2.3}$ type power-law energy spectra of parent protons continues up to $\sim$ 1 PeV. 
The efficiency of conversion of kinetic energy of stellar winds to CRs can be as high as 10 percent implying that the young massive stars may operate as proton PeVatrons with a  dominant contribution to the flux of highest energy galactic CRs.
 \end{abstract}
\newpage

There is a consensus in the CR  community that the locally detected CR protons and nuclei up to the so-called knee, a distinct feature in the CR spectrum around  1~PeV ($10^{15} \ \rm eV$), are produced in the Milky Way. Most likely,  hundreds or more  sources  contribute to 
the flux of CRs produced  at a rate $(0.3-1) \times  10^{41} \ \rm  erg/s$  \cite{drury12}. Therefore,  the hope of revealing the galactic factories of CRs on a source-by-source basis  is not affordable, especially given that many of the CR sources are not active anymore.  A more feasible approach seems to be the search for potential source populations,   the best-studied representatives of which could  (a)  collectively provide the CR production rate in the Galaxy,  and  (b)  explain 
the composition,  energy spectrum and anisotropy of CRs.   

The current paradigm of Galactic CRs \cite{blasi13} 
is based on the conviction that the supernovae (SNe) explosions, in general, and their remnants (SNRs), in particular,  satisfy both requirements. 
Over many decades, this belief has been supported by phenomenological arguments and theoretical meditations. The recent
observations conducted  in high ($E \geq 0.1 \ \rm GeV$) and very high  (VHE;  $E \geq 0.1 \ \rm TeV$) energy bands  
confirm the effective acceleration of CRs  in SNRs. 
 %
Yet, the VHE $\gamma$-ray spectra reported from more than ten young supernova remnants,  appeared to be  steep which can be explained by the  "early" cutoffs/breaks below 10~TeV (see "Methods" and the supplementary information). This has raised doubts in the CR community regarding the 
ability of SNRs to operate as CR {\it PeVatrons}.   

Moreover, the theoretical developments of recent years revealed serious problems  in standard  schemes for boosting particles  to PeV energies.  In principle,  Type II SN shocks propagating through the dense wind of the progenitor star can accelerate particles up to 1~PeV, but the  "PeVatron phase" could be accomplished only during the first  years of the explosion  (see, e.g. ref\cite{bell13, cardillo15, zirak15}).  
In this regard, the youngest known  SNR in our
Galaxy,    SNR G1.9+0.3,  is of a great interest.  Despite its young age ($T \approx 100$~yr)
and the shock speed ($v\approx 14,000$~km/s) \cite{borkowski10},  
this source currently does not operate as a PeVatron, as it follows from the position of the 
cutoff in the spectrum of synchrotron X-rays \cite{G1.9_sync}.  One cannot exclude that the PeV protons have been accelerated at earlier epochs, but,  because of the particle escape,  the remnant is already  emptied.  At first glance,  the early acceleration and escape of PeV 
protons reduces dramatically the chances of finding these  PeVatrons \cite{cristofari18}.  
However, in young  (e.g.  Tycho and Cassiopeia A)   and especially very young (e.g. SN 1987a and SNR G1.9+0.3)  SNRs, multi-TeV particles cannot run too far away from their remnants. 
For example, in the case of G1.9+0.3 located in the
Galactic Center, the propagation depth of  $E \geq  10$~TeV protons for 100 yrs  hardly could  exceed cannot  exceed 
30~pc.
For the distance to the source of 8.5~kpc,
the angular size of this region is expected less than  10 arcmin, therefore the upper limit on the  
 $\gamma$-ray luminosity $L_\gamma(\geq 1 \ \rm TeV) \leq 2 \times 10^{32} \ \rm erg/s$  reported by the H.E.S.S collaboration\cite{hessG1.9} can be applied to the total 
 energy of CRs  contained within $R \leq 30$~pc environment around the source:
\begin{equation}
W_{\rm p}(\geq 10 E) =L_\gamma(\geq E) \  t_{\pi^0} \  \eta^{-1} ,
\label{W-L}
\end{equation}

where $t_{\pi^0} \simeq 1.5 \times 10^{15} n^{-1} \  \rm s$
is the radiative cooling time of protons through the   $\pi^0$  production and decay channel;  the parameter $\eta \approx 1.5-2$  takes into account the production of $\gamma$-rays  in interactions with involvement of nuclei of both CRs and ISM\cite{kafexhiu14}.    The gas density in the region surrounding 
G1.9+0.3 is  very  high, $n \simeq 100 \ \rm cm^{-3}$, 
allowing us to constrain the  energy content 
of 0.01-1 PeV protons by $\approx  10^{45}$~erg, i.e. several  orders of magnitude  below the "nominal"   CR 
release in a SNR. 

Over the last decade, the space- and ground-based telescopes have revealed many classes of galactic $\gamma$-ray source populations. Some of them can be considered as complementary or alternative (to SNRs)  CR factories.  The clusters of young stars are of particular interest. The interacting winds of massive stars have been recognized as potential CR accelerators as early as in the 1980s. The acceleration could take place in the vicinity of the stars \cite{casse80, cesarsky83} or in superbubbles, multi-parsec structures caused by the collective activity of massive stars\cite{bykov2014, parizot04}.  
The acceleration of multiple shocks can raise the maximum energy of CR protons out of 1~PeV\cite{bykov82,Klepach2000} which makes the stellar clusters attractive candidates for cosmic PeVatrons.  

The young star clusters contain sufficient kinetic energy, supplied by interacting stellar winds, 
the conversion of which to CRs  might be traced by  secondary $\gamma$-rays, the products of  interactions of CRs with the surrounding gas. 
The diffuse  GeV  $\gamma$-rays  detected by {\it Fermi} LAT telescope around 
the compact clusters  Cygnus OB2\cite{fermi_cygnus},  NGC 3603\cite{yang17} 
and Westerlund~2 \cite{yang18} can be  naturally interpreted within this  scenario.  
%
Spectroscopic and morphological studies of moderately extended sources can be best performed at TeV energies with the atmospheric Cherenkov telescope arrays\cite{ABK08}.   
Diffuse  TeV $\gamma$-ray structures  have been indeed reported  
by the H.E.S.S  collaboration in the vicinity of the stellar cluster 
Westerlund~1\cite{hess_w1}  as well as  
30 Dor C\cite{hess_lmc} located in the Large Magellanic Cloud.  
In the case of Cygnus Cocoon,  multi-TeV $\gamma$-rays have been reported as a smooth continuation of the GeV $\gamma$-ray spectrum up to $\sim 10$~TeV  \cite{Cyg_argo}. 

The $\gamma$-ray morphology  combined with  
measurements  of the atomic and molecular gas,  
can serve as a  powerful tool  for revealing the locations and the regime 
of operation of CR accelerators \cite{aa96}.
The method  has been successfully applied 
to the diffuse  TeV $\gamma$-ray emission of the Central Molecular Zone (CMZ) 
in the Galactic Centre  (GC)\cite{hess_gc16}.  While the hard spectra of $\gamma$-rays  extending to energies of  tens of TeV,  indicate the presence of a proton PeVatron(s) in CMZ, the $1/r$ type radial distribution of parents protons up to $\sim200$~pc, points  to the continuous 
 operation of  proton PeVatron(s)  located within the central 10~pc of GC.   The supermassive black hole in the GC  has been suggested as a potential source of PeV protons\cite{hess_gc16}.   Below  we argue that  the  compact  stellar clusters, 
{\it Arches}, {\it Quintuplet}  and  {\it Nuclear} in GC, 
could  be alternative sites for the CR acceleration. To support  the hypothesis of association of the diffuse multi-TeV $\gamma$-ray emission of CMZ 
to the stellar clusters in GC, we explored the possibility of extraction of spatial distributions of CRs in the proximity of other clusters embedded in diffuse $\gamma$-ray structures.

\section{Results}
%

%
For  Cygnus Cocoon, we analysed  {\it Fermi} LAT data using the standard LAT  software package.  For  the TeV source HESS J1646-458 linked to Westerlund~1,  we used the radial profiles published by the H.E.S.S collaboration\cite{hess_w1}.  For the distribution of molecular hydrogen, we applied the data from the  CO galactic survey performed by the  CfA 1.2m millimetre-wave Telescope, while for the atomic hydrogen  we used the data from the  Leiden/Argentine/Bonn (LAB) Survey.

\begin{figure}
\centering
\includegraphics[width=0.5\linewidth,angle=-90]{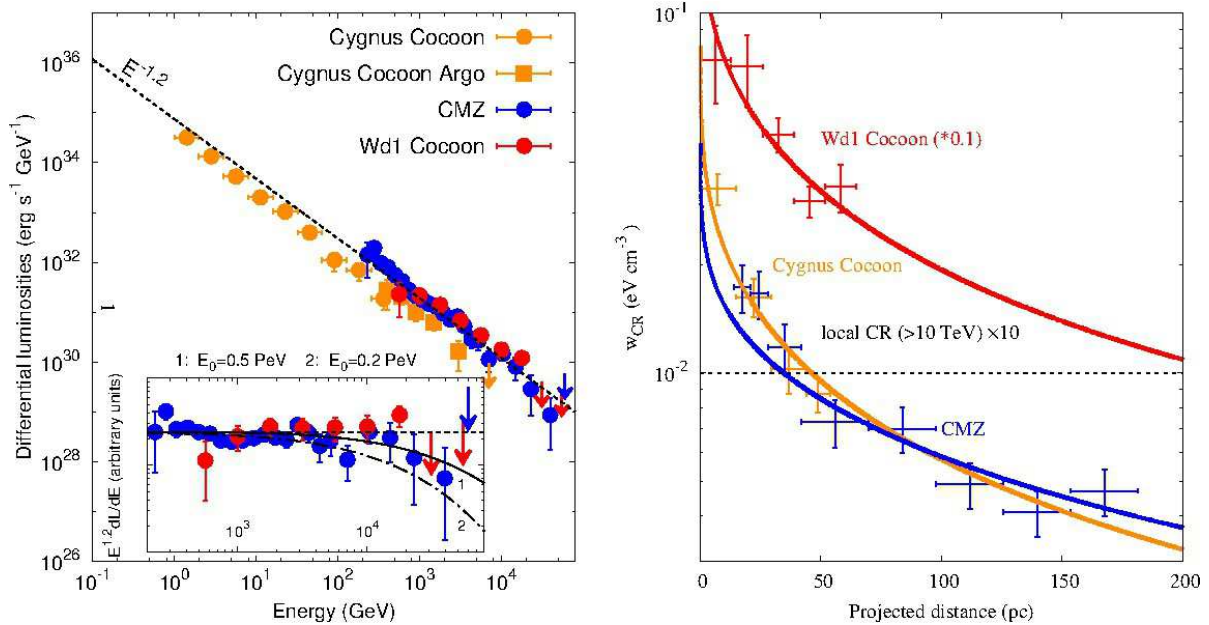}

\caption{Gamma-ray  luminosities and the radial distributions of CR protons
in extended regions  around 
the star clusters Cyg OB2  (Cygnus Cocoon) and Westerlund~1 (Wd~1 Cocoon), as well as in   the Central Molecular Zone 
(CMZ)   of the  Galactic Centre 
assuming that  CMZ is powered by CRs  accelerated in 
{\it Arches}, {\it Quintuplet}  and  {\it Nuclear} clusters. The error bars contain both the statistical and systematic errors. 
{\bf (a)} {\it  left panel}: The differential $\gamma$-ray luminosities, 
${\rm d}L/{\rm d} E= 4 \pi  d^2 E f(E)$. 
The luminosities of all three sources have similar energy dependences close to 
$E^{-1.2}$ as is illustrated by the dashed line. The inserted figure shows the differential luminosities of CMZ and Wd 1 multiplied by  $E^{1.2}$ for  a clearer illustration 
of the spectra at highest energies. We show also the gamma-ray spectra expected from interactions of parent  proton population with a spectrum of $E^{-2.3}\rm{exp}(-{E/E_0})$, with $E_0 = $ 0.2 PeV and 0.5 PeV, respectively.  
{\bf (b)} {\it right panel}:   The CR proton radial distributions  
in Cyg Cocoon, Wd~1 Cocoon and  CMZ above 10~TeV. For the Cygnus Cocoon, the energy density of protons above 10 TeV  is derived from the extrapolation of  the Fermi LAT  gamma-ray data to higher energies. The  flux reported by the ARGO collaboration at 1 TeV supports the  validity of this extrapolation.
The $\gamma$-ray flux enhancement factor 
due to the contribution of  CR nuclei is assumed $\eta=1.5$.
For comparison,  the energy densities of CR protons 
 above 10 TeV based on  the  measurements by AMS  
 are also shown  \cite{ams02proton}.  
 }
\label{main}
\end{figure}

The main conclusion following from the results presented in Methods section  is that the CR density declines as $r^{-1}$ up to $\approx$50 pc from both stellar clusters. 
The results are shown in Fig.\ref{main}b, together with the earlier published radial distributions of CR protons in CMZ  \cite{hess_gc16}. 
We show the differential $\gamma$-ray luminosities of  extended sources associated with Cyg OB2, Westerlund~1 and CMZ. The energy distributions of $\gamma$-rays are quite similar; $dN/dE \propto E^{-\Gamma}$ type differential energy spectra with 
power-law index $\Gamma \approx 2.2$ extend to 10 TeV and beyond without an indication of a break.  The $\gamma$-rays are likely to originate from  interactions of CRs with the ambient gas through the production and decay of neutral $\pi$-mesons (see below).   Because of the increase of the $\pi^0$-meson production cross-section with energy, the spectrum of secondary $\gamma$-rays is slightly harder compared to the spectrum of parent protons, $\Gamma \approx \alpha_{\rm p} - 0.1$~\cite{kelner}, thus the power-law index of the proton distribution should be
$\alpha_{\rm p} \approx 2.3$. 

\begin{table}
\begin{tabular}{|c|ccc|}
 \hline
Source&  Cyg Cocoon & CMZ  & Wd~1 Cocoon \\
 \hline
 \hline
Extension (pc) &  50 & 175 & 60\\
Age of cluster (Myr)\cite{figer08} &  3--6 & 2--7 & 4--6  \\
L$_{kin}$ of cluster (erg/s) & 2$\times$10$^{38}$~\cite{fermi_cygnus}    & 1$\times$10$^{39}$~\cite{muno06} & 1$\times$10$^{39}$~\cite{hussmann13}\\
Distance (kpc) &  1.4 & 8.5 & 4 \\
$\omega_o(> 10 \rm TeV)$ (eV/cm$^3$) &  0.05    & 0.07 & 1.2 \\
\hline
\end{tabular}
\caption{Physical parameters of  three extended $\gamma$-ray structures and the related stellar clusters}
\end{table}

The apparent similarity of the radial  ($\propto r^{-1}$) and energy ($\propto E^{-2.3}$) distributions of CR protons for 
different stellar clusters  is a hint that we observe the same phenomenon.  
The  most natural explanation  of the  $1/r$ dependence of CR radial distribution is that relativistic particles have been continuously injected into and diffused away in the interstellar medium (ISM). 
Given the tight energy budget, the diffusion time cannot be much shorter than the age of the stellar cluster (see below), $t \geq  10^6$~yr.  On the other hand, the acceleration of  multi-TeV CRs in an individual  SNR cannot last much more than  $10^3$~yr (see, e.g., ref\cite{bell13}). Thus,  to support the quasi-continuous CR injection, an unrealistically high  rate of $\sim1$ SN per 1000~yr in the cluster is required.  In our view, this disfavors SNRs and gives  preference to massive stellar winds as particle accelerators.

In the case of spherically symmetric 
diffusion,  the CR density at a distance from  the central source  $r$  depends on the injection rate $\dot{Q}(E)$ and  the diffusion coefficient:  $w(E,r) \propto \dot{Q}(E)/r D(E)$, i.e.,   the $1/r$  profile is independent of the absolute value of the diffusion coefficient unless the latter varies dramatically over the scales of tens of parsecs.  Generally,
the diffusion  coefficient is a function of energy.   Typically  it is presented in the form 
$D(E)=D_0 (E/E_0)^\delta$. We normalise the diffusion coefficient at $E_0=10$~TeV. 
Inside the accelerators and in the vicinity of powerful objects, 
the diffusion coefficient can significantly deviate from the average interstellar value.  Because of the energy-dependence of  diffusion coefficient,  
the initial acceleration spectrum,  $Q(E) \propto E^{-\alpha_0}$, outside the source is
modified,  ${\rm d}N/{\rm d}E \propto \dot{Q}(E)/D(E) \propto E^{-(\alpha_0+k\delta)}$, with $k=3/2$
for the burst-type  injection, and $k=1$ for a continuous source \cite{AAV95}.

The above relations are valid when the energy losses of CRs can be neglected. While for CR protons and nuclei this is a  fully justified assumption, relativistic electrons undergo severe energy losses.  However, 
electrons cannot be responsible for the observed   $\gamma$-ray images.     
The leptonic  (inverse Compton; IC) origin of $\gamma$-rays is excluded both at GeV and TeV energies.  Firstly,  the propagation of multi-TeV ($\gg 10$~TeV)  electrons in the ISM could hardly exceed 100 pc \cite{AAV95}.  Moreover, 
inside  a typical  cluster of a radius less than 3~pc  and  the overall starlight luminosity of  $L_{\rm r} \approx 10^{40} \ \rm erg/s$, 
the energy density of optical photons exceeds   
$u_{\rm r} \sim L/4  \pi r^2 c  \approx  100 \  \rm eV/cm^3$.  Outside the cluster,  $u_{\rm r}$  decreases as  $1/r^2$,  thus, up to tens of parsecs,  it dominates over the average radiation density in the Galactic plane.  Therefore,  in the case of IC origin of $\gamma$-rays,  we would expect a sharp increase of the $\gamma$- ray intensity towards a bright central source coinciding with the cluster.  The brightness distributions of the observed $\gamma$-ray images  of objects discussed in this paper do not agree with this prediction.

It is convenient to write the radial distribution of the CR density in the form
\begin{equation}
 w(r)=  w_0 (r/r_0)^{-1} \ .
 \label{w(r)}
\end{equation}
Below we will adopt $r_0=10$~pc, i.e. normalise the CR proton density $w_0$ 
outside but not  far from the cluster.  
The directly derived  values of $w_0(\geq 10 \  \rm TeV)$   for the Cygnus Cocoon, CMZ, and  Wd1 cocoon,   are presented in Table~1.
 
For  the radial distribution of CRs, given in the form of Eq.(\ref{w(r)}),  the total energy of  CR protons  within the volume of the radius  $R_0$  is  
\begin{equation} 
W_{\rm  p}= 4 \pi \int_0^{R_{\rm 0}}  w(r) r^2  \,\mathrm{d}r  \approx \\
 2.7 \times 10^{47} (w_0/1 \ \rm eV/cm^3) (R_0/10 \ \rm pc)^2 \ \rm erg \ .
 \label{Wcr}
 \end{equation} 
 For given $w_0$,  the  main 
uncertainty in this estimate is related to the upper limit of integration, $R_0$.  If we use for $R_0$ the distances  $R_{\rm obs}$ corresponding to the last points in   Fig.\ref{main}b,  and the values of  $w_0(\geq 10 \  \rm TeV)$  from Table~1, for protons with energy exceeding 10~TeV 
we obtain $W_{\rm p} \approx   
 3.4 \times 10^{47}, \  4.7  \times 10^{47}, \ 8.1  \times 10^{48}$  erg  for 
Cygnus Cocoon, CMZ,  and  Westerlund~1 Cocoon,  respectively.  Note that these estimates are less biased  compared to the approach  based on 
Eq.(\ref{W-L}).  Indeed,  the latter gives information only about the protons contained in the 
volumes  visible in $\gamma$-rays, thus "misses" the protons contained in  low density 
(not detectable in $\gamma$-rays) gas regions.  
This estimate strongly depends  on the  value of 
$R_{\rm obs}$ which is determined by the brightness of the $\gamma$-ray image. 
The extensions of  the  large diffuse structure  depend  
on the detector's  performance,  the level  of the  background,  {\it etc. }
Thus,   the content of CR protons within $R_{\rm obs}$  does not provide information about {\it all} CRs injected into ISM.  
The latter can be calculated by integrating  Eq.(\ref{w(r)}) up to the so-called diffusion radius $R_{\rm D}$, 
 the maximum distance penetrated by a particle of energy $E$ during the time $T_0$.  In the case of   negligible energy loses of propagating particles, 
\begin{equation}
R_{\rm D}  =2 \sqrt{T_0 D(E)}  \approx   3.6  \times 10^3  (D_{30}  T_{6})^{1/2} \  \ \rm pc,
 \label{R_D}
 \end{equation}
where $D_{30}$  is the diffusion coefficient  of protons in  the units of  $10^{30} \rm cm^2/s$, 
and $T_{6}$ is  $T_0$ normalised to $10^6$ years. The ages of the 
individual clusters  vary  in a narrow range  between 2  and  7~Myr (see Table 1). 
In the source neighborhood,  the diffusion coefficient   
cannot be very large  otherwise the 
demand on the total energy in CRs would  exceed  
the available energy contained in the stellar winds,
\begin{equation}
W_{\rm  tot}=f L_0 T_0=3 \times 10^{52}f  L_{39} T_6 \ \rm erg, 
 \label{W_CR}
 \end{equation}
where $L_{39}=10^{39} L_0$ is the 
total mechanical power  of the stellar winds  in units of $10^{39} \ \rm erg/s$, 
and $f$ is the efficiency of conversion of the 
wind kinetic energy to relativistic protons with energy larger than 10 TeV.   
Substituting $R_0=R_{\rm D}$ into Eq.(\ref{Wcr}), we obtain
\begin{equation}
f (\geq 10 \ {\rm TeV})  \approx   1 w_0 D_{30}  L_{39}^{-1}  ,
 \label{f_D} 
 \end{equation}

Eq.(\ref{f_D}) provides a direct link between two fundamental parameters characterizing the efficiency of acceleration of CRs inside the stellar cluster and their propagation  in the $\sim 100$~pc vicinity of the accelerator.  
At 10 TeV, the diffusion coefficient in the Galactic Disk is estimated $D_{30} \sim 1$ \cite{strong07}.  Closer  to the powerful particle accelerator,
the CR diffusion could be slower because of  higher turbulence  
maintained, in particular,  by  CRs. If the linear size of the 
observed $\gamma$-ray image $R_0=60$~pc  reflects  the propagation depth of CRs, $R_{\rm D}$,   
Eq.(\ref{R_D})  gives  $D \sim  5 \times 10^{25} \ \rm cm^2/s$.   
For the interstellar magnetic field $B \sim 10 \mu$G,  the latter  
is comparable to the  Bohm diffusion coefficient, 
$D_{\rm B}=R_{\rm L} c/3 \approx 3 \times 10^{25} 
(E/10 \rm TeV) (B/10 \mu \rm G)^{-1} \ \rm cm^2/s$. It seems quite 
unrealistic, implying that the radius of the CR halo around Westerlund~1 most likely is significantly larger than the angular size of the resolved $\gamma$-ray image.

Eq.(\ref{f_D})  gives an alternative  estimate for the diffusion coefficient.   The overall wind power of massive stars in Westerlund~1 is estimated $10^{39} \  \rm erg/s$, while the required energy density of $\geq 10$ TeV protons  is $w_0=1.2 \ \rm eV/cm^3$ (see Table 1).
Assuming that the efficiency of acceleration of  $f (\geq 10 \ \rm TeV)$ protons cannot  be significantly larger than  0.01 (given that
for the spectrum $E^{-2.3}$  the   content  of $\geq 10$~TeV protons
does not exceed 10 \% of  the overall CR energy above 1 GeV),
we can estimate the  upper limit on the diffusion coefficient around 10 TeV.  From Eq.(\ref{f_D}) follows that $D$
should be smaller, at least by a factor of 100, compared to the diffusion coefficient in the ISM, i.e., $D \sim 10^{28} \ \rm cm^2/s$.

For this value,  the size  of the expanding cloud of CRs can be as large as 300~pc
(see Eq.(\ref{R_D}).  Remarkably, from Fig.1b follows that
even at such large distances  the density of $\geq 10$~TeV
CRs could exceed the locally measured  CR density by two orders of magnitude. Thus one should expect an extension of enhanced $\gamma$-ray emission well beyond
the angular size of $\approx 1^\circ$  \cite{hess_w1},
unless the gas density
in this region is not anomalously low.   The same is true for the Galactic Centre and the
Cygnus region (see Fig. 1b). Detailed spectrometric and spatial studies of diffuse multi-TeV structures
extending up to several degrees,  can be provided by the Cherenkov Telescope Array (CTA) \cite{CTA}. Because of the cutoff in the CR spatial distribution at $R_{\rm D}$,  one should expect relatively sharp edges in the corresponding $\gamma$-ray images.
For the given age of the cluster,   the identification of the diffusion radius
would  give unambiguous information  about the diffusion coefficient and, consequently, about  the acceleration efficiency $f$ through
Eq.(\ref{R_D}) and  Eq.(\ref{f_D}),  respectively.  Regarding the observational perspectives,  one should note that the chance probability of appearance of background or foreground structures in the extended gamma-ray images surrounding stellar clusters, is quite high. In the Galactic Disk, these diffuse structures can be related to unresolved sources belong the largest source populations - SNRs, Pulsar Wind Nebulae and Giant Molecular clouds.  Such a problem we face, most likely, in the case of Cygnus and Wd1 Cocoons. 

The  spectra of  CR protons  inside  of all three  diffuse $\gamma$-ray sources 
are  described by power-law energy  distributions with an index 
$\alpha_{\rm p}  \approx 2.3$. 
It is formed from the initial (acceleration) spectrum of protons, 
$\dot{Q}(E) \propto E^{-\alpha_0}$,  but can be modified due to the energy dependent diffusion, $J_{\rm p}(E, r) \propto \dot{Q}(E)/D(E) r^{-1}$.  For the Kolmogorov  type turbulence, $D(E) \propto E^{1/3}$, we  arrive at  a "classical"  $E^{-2}$ type acceleration spectrum. One cannot, however,  exclude that at energies  $\geq 10$~TeV  the diffusion slightly 
depends on the energy. Even in this extreme case the 
acceleration spectrum of protons would be relatively hard with   
$\alpha_0=\alpha_{\rm p} \approx 2.3$. 
The hard $\gamma$-ray spectra of Westerlund 1 Cocoon and CMZ continue 
up to 20-30 TeV without  an indication of a cutoff or a break.  Correspondingly, the energy spectra of  parent protons should  not break 
at least until 0.5~PeV (see Fig.\ref{main}a).   This makes the clusters of massive stars potential sources of multi-TeV neutrinos with a fair chance to be detected by the cubic-km volume neutrino detectors.  In particular, 
Westerlund~1, which has the highest $\gamma$-ray flux at 20 TeV, 
seems to be  a promising target for neutrino observations \cite{Bykov_nu}.

\section{Conclusions}

The stellar clusters offer a viable solution to the long-standing problem of the origin of  Galactic CRs with  massive/luminous stars as major contributors to observed fluxes of CRs up to the {\it knee} around 1 PeV.   In the context of available energetics and the acceleration 
efficiency, the population of young stellar clusters and SNRs are an equally  good choice. The same is true for the  speeds of outflows (stellar winds and SNR shocks) of several thousand
km/s, which is a key condition for the effective utilization of the diffusive shock acceleration mechanism in the PeVatron regime.   

Yet, we argue that the stellar clusters have a certain advantage compared to SNRs as long as it concerns the multi-TeV to PeV domain. Undoubtedly, the hard power-law spectra of $\gamma$-rays from extended regions  surrounding the stellar clusters  are of hadronic origin. The extension of these spectra beyond 10~TeV without an indication of a break  points
out that the massive stars can operate as PeVatrons. Remarkably, the potential of stellar winds (and their advantages compared to SNRs!) to accelerate protons to PeV energies, has been foreseen by Cesarsky and Montmerle \cite{cesarsky83} as early as 1983! .   

In contrary, the spectra of SNRs including the prominent representatives like Tycho, Cassiopeia A and SN 1006,  are steep or contain breaks at energies below 10~TeV.  As a result,  a suspicion is mounting among the experts that SNRs do not operate as PeVatrons,  and, thus, this source population alone cannot be responsible for the overall flux of galactic CRs. This conclusion does not concern the lower energy band, where SNRs can make a major contribution.  Interestingly, if both source populations convert the available mechanical energies of the stellar winds and SN shocks to CRs with similar (10 \% or so) efficiencies,  but different acceleration spectra
($\alpha_0 \leq 2.3$ for stellar clusters and $\alpha_0 \geq 2.4$ for SNRs), one should expect a dominance of SNRs in the sub-TeV and the stellar clusters in multi-TeV bands of the spectrum of Galactic CRs. In this case, one may expect a spectral change in the transition region which can be a possible explanation of the  hardening observed directly of the CR spectrum above 200 GeV \cite{ams02proton}. Note that both source populations are in a good agreement with the 
conclusion derived from the reported content of heavy isotopes, that the CR  acceleration in our Galaxy takes place in regions populated 
by massive OB stars and  SN explosions\cite{Ellison,binns16}.

To conclude, the multi-TeV $\gamma$-ray observations provide evidence that the clusters of massive stars operating as PeVatrons may substantially contributing to the flux of galactic CRs. The extension of spectrometric and morphological  $\gamma$-ray measurements up to 100 TeV in the energy spectrum and up to several degrees in the angular size,  from regions surrounding powerful stellar clusters would provide crucial information about the origin of CRs in general, and the physics of proton PeVatrons, in particular. Such observations with the Cherenkov Telescope Array will be available in coming years.

\begin{methods}

\section{Fermi data analysis of Cygnus Cocoon}
For the analysis of {\it Fermi} LAT data, we have selected observations
towards the Cygnus region  for a period of more than 9 years (MET 239557417 -- MET 532003684),
and used the standard LAT analysis software package \emph{v10r0p5}(\url{http://fermi.gsfc.nasa.gov/ssc}) The   $P8\_R2\_v6$ version of the  post-launch instrument response functions (IRFs) was used, and both the front and back converted photons were selected.

For the region-of-interest (ROI),  a $15^{\circ} \times 15^{\circ}$ square area centred on the point  of  $RA_{\rm J2000}=307.17^{\circ}$, $DEC_{\rm J2000}=41.17^{\circ}$ 
has been chosen. The observations with "rock angles"  larger than $52^{\circ}$ were excluded. 
In order to reduce the effect of the Earth albedo background, we also excluded  the time intervals when  the parts of the ROI were observed at zenith angles $> 90^{\circ}$. %
Also, for the spatial analysis, given the crowded nature of the region and the large systematic errors due to a poor angular resolution at low energies, we selected only  photons with energies exceeding 10~GeV at which the angular resolution is significanty  improved achieving to $\sim0.1^{\circ}$.
Note that this energy cut dramatically reduces the possible contribution of  pulsars which 
are bright only at energies below a few GeV.

The $\gamma$-ray count map above 10 GeV  in  the  $10^{\circ} \times 10^{\circ} $ region around  Cygnus Cocoon is shown on the  left panel of  Supplementary Figure.1. We performed a binned likelihood analysis by using the tool {\it gtlike}.  The point sources listed in the 3rd Fermi source catalog (3FGL) \cite{3fgl} are also shown; the identified sources are shown with blue crosses,
while  the red crosses  indicate the positions of non-identified objects in the 3FGL  catalog.  
We also added  the background models provided by the Fermi collaboration (gll\_iem\_v06.fits and  \_P8R2\_SOURCE\_V6\_v06.txt  for the galactic and the isotropic diffuse  components, respectively( available 
at \url{http://fermi.gsfc.nasa.gov/ssc/data/access/lat/BackgroundModels.html}). In the analysis, the normalisations and the spectral indices of sources inside the FOV were left free.  We used the 2-D gaussian template provided by Fermi Collaboration to model the extended emission from 
the Cygnus Cocoon.   We varied the position and the radius of the Cygnus Cocoon template but did not find a significant  improvement.  Therefore, for derivation of the energy spectrum  we use the 2D gaussian template provided by the {\it Fermi} LAT Collaboration.   In the ROI  two TeV sources are also detected \cite{abdo12},  Gamma Cygni and TeV J2032+415. We note that both of them are already identified in the 3FGL catalog and included in this analysis.  The position of both sources are shown in the right panel of Supplementary Figure 1.

To derive the spectrum we divided the energy interval $0.5~ {\rm GeV} - 500~{\rm GeV}$ into 10 logarithmically spaced bands and applied the tools \emph{gtlike} to each of these bands.    The results
are consistent with the results reported in \cite{fermi_cygnus}. The larger photon statistics  and the new data reduction tools allow significant extension of the spectrum, 
up to 500~GeV.  In the following study both statistical and systematic errors (due to the systematic uncertainties of the effective area \cite{3fgl}) are included in the following analysis. 
The spectrum above 1 GeV is well fitted with a power law with a photon index of $2.2 \pm 0.1$ and  integrated flux of  $1.0 \pm 0.1~\rm \times 10^{-7} ph ~cm^{-2}~ s^{-1}$.  The detected spectrum extends to 500 GeV without an indication of a cut-off or a break.  
Correspondingly, the parent  protons should have  a power-law spectrum with a slightly 
larger ($\approx 0.1$ spectral index  up to $\approx 20 \times 0.5 \ \rm TeV=10 \ \rm TeV$)
\cite{kelner}.   

We should note that multi-TeV $\gamma$-rays  have been 
claimed from the Cygnus Cocoon 
by  the MILAGRO \cite{abdo12} and  ARGO \cite{Cyg_argo} 
collaborations.   
The comparison of spectral points from different experiments  requires 
a special and non trivial treatment given the statistical and systematic uncertainties 
concerning the energy measurements, as well as the different extensions of the regions from which the $\gamma$-rays have been detected.  Nevertheless, for the sensitivities of these detectors, the reported fluxes  could hardly appear below the extrapolation of the {\it Fermi LAT} spectrum of Cygnus Cocoon.


\section{Radial distribution of Cosmic rays}
 The brightness distribution of $\gamma$-rays is shaped by  the  product 
of spatial distributions of  CRs and the gas density.  In the ISM,  
the dense gas  complexes are  distributed rather chaotically. 
Therefore,   the probability of 
detection of  an "ordered"   $\gamma$-ray image is small.  
In particular, the $1/r$ type smooth radial distributions of CRs  originating from 
the young star  clusters Cyg OB2 is derived from quite  irregular $\gamma$-ray images.  The position of the 
star cluster Cyg OB2  is  significantly shifted from the centre of the 
surrounding $\gamma$-ray image  (see Supplementary Figure 1).

The comparison of  the spatial distributions of the $\gamma$-ray brightness derived in the previous sections  and the 
gas density derived in the supplementary informations
does not show linear correlation  which one would expect  in the case 
of homogeneously distributed parent CRs.  
To investigated the  CR distribution, we produce the radial profile of the the $\gamma$-ray emissivity,
which is proportional to the CR density.

\subsection{Cygnus Cocoon}
 As the reference point we take the position  
 of the  stellar cluster Cygnus OB2.  As noted above, Cygnus OB2 is  not symmetrically located inside the 
 $\gamma$-ray image of the Cygnus Cocoon. Its choice as the reference point is motivated by the 
 hypothesis that the massive OB stars of this cluster are the main 
 producers of CRs responsible for the $\gamma$-ray emission.

The $\gamma$-ray flux is derived above 10~GeV, using the standard likelihood analysis, 
for five rings centred on  Cygnus OB2 within the following angular radii:  
 [0:0.4]$^{\circ}$, [0.4:0.8]$^{\circ}$, [0.8:1.4]$^{\circ}$ [1.4:1.8]$^{\circ}$  and [1.8:2.2]$^{\circ}$.
 We note that the presence of the bright $\gamma$-ray pulsar LAT PSR J2023+4127  close to 
 Cygnus OB2 may introduce additional contamination.  To minimise the impact of this pulsar,  we performed the so-called 
 pulsar gating analysis. Namely, we produced the phase-folded light curve of  LAT PSR J2023+4127 
 using  the  ephemeris  corresponding to the time interval  from MJD 54658 to 56611 (see \url{https://confluence.slac.stanford.edu/display/GLAMCOG/LAT+Gamma-ray+Pulsar+Timing+Models}). To obtain the light curve, we adopted a $1^{\circ}$ aperture without applying any background subtraction above 1 GeV.  The resulted light curve is shown in Supplementary Figure 6.  It shows two peaks located at the phases 0.5 and 0.95, respectively. Therefore,  in order to remove the pulsed emission,  the $\gamma$-ray data have been 
 selected only for  the phase intervals  [0.1,0.4] and [0.6,0.9].  In this way, we dramatically 
 reduce the impact of  the bright $\gamma$-ray pulsar, albeit  at the  expense of reduction  of the
 $\gamma$-ray  photon statistics by 40\%.    

The total gas column density  is contributed by the  molecular, neutral atomic and ionised 
hydrogen components as discussed in the supplementary informations.  The comparison of 
spatial distributions of  the $\gamma$-ray intensity and the overall gas column density
gives the radial profile of $\gamma$-ray emissivity, and consequently provides direct 
information about the profile of the CR density. 

The derived radial profile of the $\gamma$-ray emissivity 
is shown in  Supplementary Figure 7 together with two  curves corresponding  to (i)   the 
the homogeneous distribution of CRs which is formed in the case of an 
impulsive injection event, and  (ii)   $1/r$ type distribution of CRs, which is expected in the case of 
continuous injection of CRs  into ISM. The latter 
distribution is favoured, with a $\chi^2$/ndf of 0.99 versus a $\chi^2$/ndf of 31.0,
compared to the  case of homogeneous distribution of CRs. 
 Due to the projection effect, the curves shown in Supplementary Figure 7  have the  form $f(r)\sim log((r_0+\sqrt{r_0^2-r^2})/r)/\sqrt{r_0^2-r^2}$, where $r$ is the radial distance and $r_0$ is the size of the emission region. Here we assume   a spherical symmetry for  the $\gamma$-ray emission region and integrate the density in the line of sight inside the region.

 We note that the TeV source TeV J2032+415 \cite{atel10971}  coincides in position with the pulsar LAT PSR J2023+4127 and may contaminate the most inner bin. This source has been recently identified as a variable gamma-ray source, with a flux level (above 200 GeV) increased by a factor of 2 in a recent flare event \cite{atel10971}. However, even without the first bin the  $1/r$ type distribution of CRs are  strongly favored, the $\chi^2$/ndf  is 0.8 compared with a  $\chi^2$/ndf of 15.6 in the homogeneous CR distribution case.
 
 To test the possible azimuthal variation of the profile we also divided the rings into south and north hemisphere  and derive the radial profile therein, respectively.  The derived profile are plotted in Supplementary Figure 2.  We note that due to limited statistics for each hemisphere we have three bins. The profile are also consistent with the 1/r profile with the error bars. 

\subsection{Westerlund 1 Cocoon} 

We use the H.E.S.S published results for the radial profile of the VHE $\gamma$-ray excess \cite{hess_w1}. Due to the limited angular resolution of the gas maps we rebinned the H.E.S.S radial profile to a binsize of $0.2 ^{\circ}$.  We show the H.E.S.S excess map in Supplementary Figure 4. The figure is derived using the public data stored in the H.E.S.S website (\url{https://www.mpi-hd.mpg.de/hfm/HESS/pages/publications/auxiliary/AA537\_A114.html}). To minimize the contamination from the nearby TeV source HESS J1640-465/J1641-463 we omit all points beyond $1.0 ^{\circ}$. The determination of gas mass is described in supplementary informations. The results of the radial distribution of $\gamma$-ray emissivities (per H-atom) are shown in  Supplementary Figure 8. As in the case for Cygnus Cocoon, $1/r$ type distribution of CRs are  strongly favored, the $\chi^2$/ndf  is 0.9 compared with a  $\chi^2$/ndf of 11.2 in the homogeneous CR distribution case.

\subsection{Central molecular zone (CMZ)}

The $\gamma$-ray observations towards  CMZ,  as well as the corresponding gas distribution have been studied comprehensively  in the H.E.S.S collaboration paper\cite{hess_gc16}.  Here  we use the  radial profiles obtained in that work.

\subsection{Normalized emissivity maps}   The gamma-ray emissivity per H atom, i.e., normalized to the gas density, depends linearly on the CR density. Thus the normalized emissivity map contains direct information about the spatial 2D distribution of the CR density.  An ideal plan would be the one with statistically significant signal in pixels with a size comparable to the angular resolution of the gamma-ray detector.  Unfortunately, because of lack of adequate photon statistics,  the production of statistically significant maps with a meaningful grid cell currently is impossible both at GeV and TeV energies.  In Supplementary Figure 9 we show the emissivity maps of the  Wd 1 and Cygnus Cocoons. The normalized emissivity maps are derived by dividing the  Wd 1 HESS excess map and the Cygnus Cocoon Fermi LAT residual map by the corresponding gas maps, respectively.  Because of the limited photon statistics,  the derived emissivity maps are dominated by Poisson noises. Indeed, For Cygnus Cocoon, the analysis of the Fermi LAT data gives a statistically significant signal  $\approx 21~\sigma$ integrated over the source occupying   $20~ \rm deg^2$. For the grid cell size of $0.3^\circ$, this implies that on average the photon statistics in each cell is less than half percent of the overall statistics. Correspondingly, on average, the significance of the signal in each cell cannot exceed $1 \sigma$. The same is true for the Wd1 Cocoon. In contrast,  the ring templates do reveal statistically significant detection in all chosen rings;  see Supplementary table 1 and table 2.   In these tables, we also show the gamma-ray luminosities, the photon indices,  the total mass of the gas and the derived CR density in the rings in the Cygnus and Wd1 Cocoons. 
 It should be noted that the CR distribution from continuous sources follows the exact $1/r$ type contribution only in the case of negligible changes of the diffusion coefficient over the tens of parsecs. Otherwise one may expect a deviation from the $1/r$ dependence as well as anomalies in certain directions.  The presence of such anomalies in data, however, should not necessarily be interpreted as a result of variation of the diffusion coefficient.  Because of the large, up to several degrees, extensions,   the gamma-ray images of the cocoons around the stellar clusters can be contaminated by random diffuse structures. The reason could the foreground or background sources,  e.g., Supernova Remnants, Giant Molecular Clouds and Pulsar Wind Nebulae,  especially in the crowded regions of the Galactic Disk, This seems the case of both Cygnus Cocoon and the Wd1 Cocoon.  One can see this in the emissivity maps shown by red boxes in Supplementary Figure 9. For Westerlund 1 the excess region occupies about 18\% of the total area of the source. The excess gamma-ray flux is about   10\% of the total emission corresponding to  30 mCrab at 1 TeV. For Cygnus Cocoon, the excess region occupies 19\% of the source  and the integrated  flux above 1 GeV is $1.5\times10^{-8}~\rm ph~cm^{-2}~s^{-1}$.  This emission could be related to the wind nebula of the pulsar    LAT PSR J2023+4127.  For the distance to the pulsar of 1.3~kpc, the gamma-ray luminosity corresponding to the the "excess" flux is about 3 \% of the spin-down luminosity of the pulsar, $L_{\rm SD} \approx 1.5 \times 10^{35} \ \rm erg/s$.
 
So far, we have included these excess regions in the data analysis. To understand the impact of these regions on the result, we re-analyzed the data for both sources but now excluding the excess areas from the treatment, and re-calculating the radial distributions of CR protons.  For Cygnus Cocoon, this is done by adding a box shaped template (labled as red box in the left  panel of Supplementary Figure 9), and redoing the likelihood fitting. We also added the emissivity map after subtracting this additional component in the bottom panel of Supplementary figure 10. After the excess has been removed,  the map shows a general symmetry around the central cluster. We fit this map with an universal $1/r$ profile ($ log((r_0+\sqrt{r_0^2-r^2})/r)/\sqrt{r_0^2-r^2}$ after projection). We then subtracted the best-fit $1/r$ profile from the emissivity map to derived the residual emissivity map. We quantify the deviation from the $1/r$ profile by dividing the residual emissivity map by the best-fit $1/r$ profile.  We found the root mean square of the deviation is 0.25.  For most pixels the deviation is below 30\%, which can be attributed to Poission noises.  

For Westerlund 1 we removed the excess region ( red box in the top right panel of  Supplementary Figure 9) and calculated the emissivities from the map directly. In both cases the "dark" gas component is taken into account. 
The results shown in Supplementary figure 10, are still consistent with the $1/r$ profile. We show the results in Supplementary figure 10. For the Cygnus Cocoon, the fitted $\chi^2$/ndf to a $1/r$ profile is 0.80, implying an improvement compared with the fitting to the whole region.  For Westerlund 1  the derived $\chi^2$/ndf is 1.1, which is slightly worse than the profile of the entire area but still provides a perfect fit.

\subsection{CR density}

To derive the CR density, we used the following expression, 
\begin{equation}
w_{\rm CR} (\geq 10 E_{\gamma}) = \frac{W_p(\ge 10 E_{\gamma})}{V} = 1.8 \times 10^{-2} \left( \frac{\eta}{1.5} \right)^{-1} \left( \frac{L_{\gamma} (\geq E_{\gamma} )}{10^{34} {\rm erg/s}} \right) \left( \frac{M}{10^6 M_{\odot}} \right)^{-1} \rm eV/cm^3 \, ,
\end{equation}
where $M$ is the mass of the  relevant region and $\eta$
accounts for the presence of nuclei heavier than hydrogen in both cosmic rays and interstellar matter. $\eta$ depends on the chemical composition of CRs and the ambient gas, and typically varied between $1.5 \sim 2.0$\cite{dermer86, mori09, kafexhiu14}. The $\gamma$-ray luminosity, mass estimates and resulted CR densities  for all regions of the  Cygnus and  Wd 1 Cocoons  are presented in Supplementary Table~1and 2. 

\section{Gas distribution}
 To evaluate the total gas around the selected clusters, we used several tracers: CO maps from the galactic survey performed by Dame et.al \cite{dame01}; HI from the Leiden/Argentine/Bonn (LAB) Survey; and dust maps from the Planck archival data base \cite{planck}. 
For the CO  data, we use the standard assumption of a linear relationship between the velocity-integrated 
CO intensity, $W_{\rm CO}$, and the column density of molecular hydrogen, N(H$_{2}$), adopting for the 
conversion factor $X_{\rm CO}=2.0 \times 10^{20}~\rm cm^{2} (K~km~s^{-1})^{-1}$ \cite{bolatto13}.
For the HI data  we use the equation 
\begin{equation}
N_{HI}(v,T_s)=-log \left(1-\frac{T_B}{T_s-T_{bg}}\right)T_sC_i\Delta v \  ,
\end{equation}
where $T_{bg}\approx2.66$~K is the brightness temperature of the cosmic microwave background radiation at 21cm, and
$C_i = 1.83 \times 10^{18} \rm  \ cm^{2}$.  For  $T_B > T_s-5~\rm K$, 
we truncate $T_B$ to $T_s-5~\rm K$; 
$T_s$ is chosen to be 150 K.

 To account for neutral gas, which may not be always traced by CO and/or HI observations (e.g. in optical thick clouds), we used an additional method provided by infrared observations. To evaluate the column density in such cases, we used the relation between the dust opacity and the column density given by the following expression (Eq.~(4) of ref.\cite{planck}, ).
 
\begin{equation}\label{eq:dust}
\tau_M(\lambda) = \left(\frac{\tau_D(\lambda)}{N_H}\right)^{dust}[N_{H{\rm I}}+2X_{CO}W_{CO}],
 \end{equation}
where $\tau_M$ is the dust opacity as a function of the wavelength  $\lambda$,  $(\tau_D/N_H)^{dust}$ is the reference dust emissivity measured in low-$N_H$ regions, $W_{CO}$ is the integrated brightness temperature of the CO emission, and $X_{CO}=N_{H_{2}}/W_{CO}$ is the  $H_2/CO$ conversion factor.
The substitution of  the latter into Eq.~(\ref{eq:dust})  gives 
\begin{equation}
N_H = N_{H{\rm I}} +2 N_{H_2} =  \tau_m(\lambda)\left[\left(\frac{\tau_D(\lambda)}{N_H}\right)^{dust}\right]^{-1}. 
\end{equation}
Here  for the dust emissivity at $353~\rm GHz$,   we use  $(\tau_D/N_H)^{dust}_{353{\rm~GHz}}=1.18\pm0.17\times10^{-26}$~cm$^2$  taken from Table~3 of  ref.\cite{planck}. 

The missing gas can be evaluated then by examining the residual maps resulting from fitting the total dust opacity map as a linear combinartion of the HI and CO maps. The fit is then iterated, including the residual map until convergence is achieved. 

This method is similar to the derivation of  the $E(B-V)_{res}$ templates used by the Fermi LAT collaboration \cite{fermi_diffuse},  
but  instead of the extinction maps we use the dust opacity maps.  Indeed, $E(B-V)$ has nearly perfect linear correlation with the dust opacity, especially in higher column regions \cite{planck13-11}. Thus our method  and the method used in ref.\cite{fermi_diffuse} should give similar results.

Note, however, that the lack of information about velocity (or distance) implies an overestimation of the gas, integrated for the whole line of sight. For fairness we show results both with and without the \emph{dark} component.

The Cygnus region is  located  inside  the Local Arm, although the Perseus and the outer arms also contribute to the total gas content in the line of sight.  To select the gas content related to Cygnus Cocoon itself,  we separate,  following  \cite{fermi_cygnus_gas}, two regions contributing to the total gas. In the $\rm H_{I}$ and CO maps, we assign the gas with $V_{\rm LSR} <-20 ~\rm km/s$ to be Local arm and  those with  $V_{\rm LSR} >-20 ~\rm km/s$ to be outer arms. The gas content related to the local arms are regarded to be connected with extended $\gamma$-ray emission in Cygnus Cocoon. The gas mass is derived in the region defined by the $\gamma$-ray 2D template.

The Cygnus region harbours   huge amount of $\rm H_{II}$ gas.  To determine the $\rm H_{II}$ column density, we use  the Planck free-free map \cite{planck15-10}.  First, we convert the emission measure (EM) in the Planck map into the free-free intensity by using the conversion factor in Table 1 of \cite{finkbeiner03}. Then,  we used Eq. (5) of \cite{sodroski97} to calculate the $\rm H_{II}$ column density from 
the intensity of free-free emission. One should  note that the derived  $\rm H_{II}$  column density is inversely proportional to the electron density $n_e$  which is  chosen here to be $2 ~\rm cm^{-3}$  \cite{sodroski97}  as a fiducial value.   Summing all gas phase the total mass amounts to $2\times 10^6 M_{\odot}$. For the radius of the Cygnus Cocoon of  70 pc, the average gas density is estimated 
between 10  to $20 \rm ~ cm^{-3}$,  given the  approximately  50\% uncertainty in  the mass estimate. We also note that the $\rm H_{II}$ component only contributes about 20\% of the total gas content, the exclusion of this component does not alter the results of the next section.

For Wd1 Cocoon we also perform a kinetic separation, the velocity range  $-60  ~\rm km/s <V_{\rm LSR} <-50 ~\rm km/s$\cite{hess_w1} for both  $\rm H_{I}$ and CO maps are chosen in this analysis. The  gas content is shown in Supplementary Figure\ref{fig:gas_w1}. The total gas mass in the TeV $\gamma$-ray emission region (about $1^{\circ}$ around Westerlund 1) is about $3\times 10^5 M_{\odot}$. 

\section{Supernova Remnants: the main contributors to Galactic  CR?}

SNRs are widely believed to be the main suppliers  of Galactic CRs.  
This conviction  is based on two sound arguments: (1) the availability of 
sufficient energy in the form of supernova explosions to support 
the required CR production rate  in the Galaxy,   and 
(2) adequate conditions in young SNRs  for  acceleration of relativistic particles through the mechanism of  Diffusive Shock Acceleration (DSA).   Yet, despite the extensive  experimental 
and theoretical studies of CRs over the last several decades,  
the SNR paradigm of the origin of galactic CRs  
should be still  confirmed.  

The direct measurements of  CRs  are important, but they cannot address the principal   
question regarding  the localisation and identification of particle accelerators.   
Therefore, the ultimate solution to this long-standing problem can be found only by astronomical means.  The acceleration of CRs in SNRs and their subsequent interactions with the ambient matter make these objects potentially detectable sources of 
$\gamma$-rays and neutrinos \cite{1994A&A...287..959D}.%
Indeed, over the last  20 years,  many  young and mid-age SNRs have been detected in 
GeV and TeV bands.  In Supplementary Figure \ref{fig:TeVSNRs}  we show the 
spectral measurements (in the form of differential luminosities) of several prominent representatives of  young SNRs at energies above 100~GeV
\cite{2014MNRAS.439.2828A,2011A&A...531A..81H,2009ApJ...692.1500A,2009ApJ...692.1500A,2007ApJ...661..236A,2007A&A...464..235A,2006A&A...449..223A,2017ApJ...836...23A,2017MNRAS.472.2956A}.  Most of them show  a  shell-like morphology  supporting the general predictions of DSA and, thus,  establishing the SNR shocks as effective particle accelerators.

The very fact of  detection of  VHE  $\gamma$-rays  does not yet 
prove  the dominant  role of SNRs in the production of galactic CRs. 
VHE $\gamma$-rays from young SNRs  demonstrate  the effective acceleration of
particles 
up to energies of 100~TeV,  but  
it is not yet clear that  the detected $\gamma$-rays are of hadronic origin.
In SNRs, in addition to the  $\gamma$-ray production
in interactions of  CR protons and nuclei with the surrounding gas, 
an equally important  process is the Inverse Compton (IC)  scattering of 
ultra-relativistic electrons on the 2.7K CMB   and infrared photons.  
For this reason,  the  origin  of  $\gamma$-radiation detected 
from {\it all} SNRs is under intense debates. The interpretations within 
the leptonic and hadronic scenarios 
have `contras' and `pros' . Within the  uncertainties of the principal parameters, 
both models can  satisfactorily fit  the broad-band $\gamma$-ray  spectra \cite{FA13}. 

`Leptonic or hadronic'  It  is one of the key issues  of  
current interpretations of   $\gamma$-ray observations of SNRs.  But it still does not address a more  fundamental question whether SNRs are the major contributors to the Galactic Cosmic Rays.  Actually, there are two questions to be addressed: 
(i)  whether SNRs can produce cosmic rays with overall energy close to $10^{50}$erg, (ii) 
if yes, whether they can be responsible for the locally observed CR flux up to the ``knee" around 1 PeV.

The positive answer to the hadronic origin of $\gamma$-radiation would not imply a positive answer to the first question as well.  Although the total energy budget in CRs  $W_{\rm p}$  derived from $\gamma$-ray data in some SNRs is close to   $10^{50}$~erg,
 in some others it is estimated significantly lower. One should notice that the estimates of $W_{\rm p}$  depends on the ambient gas density ($\propto 1/n)$ and, therefore,  contain  an order of magnitude uncertainties. 
Concerning the second question,  the term ``PeVatron''
implies an object accelerating protons with a  hard ($E^{-2}$ type)  energy spectrum without  a  break up to $E \sim 1$~PeV.   The spectrum of secondary $\gamma$-rays almost mimics  the spectrum of the parent protons but is shifted towards low energies by a factor of  20-30. Thus,  a detection  of  $\gamma$-rays with a hard-power-law energy spectrum extending several tens of TeV would imply an unambiguous detection of a PeVatron.  So far, the observations of young SNRs did not reveal such hard multi-TeV 
energy spectra.  Only a few  SNRs have been detected above 10 TeV,  but in all cases we see steep spectra,   typically with a slope  between 2.5 and 3 (see Supplementary Figure \ref{fig:TeVSNRs}). This can be interpreted either large power-law indices or \emph{early} cutoffs in the proton spectra (typically less than tens of TeV) implying that the spectra of parent protons do not extend much beyond 100 TeV.  This, to a certain  extent, unexpected result  is a hint that  either the young 
SNRs do not accelerate CR protons to PeV energies or  the production of  PeV protons takes  
place in  objects  belong a sub-class of SNRs  which so far have not been 
detected in $\gamma$-rays. The first option 
 would imply the inability of SNRs to play the major role in the production of galactic CRs. 
 The second   option leaves  a room for  "accommodation" of  SNRs in the scheme 
 of galactic CRs.  But it means that only a small fraction of SNRs contribute to  CRs, at least at 
 highest energies. Consequently, one should assume that the efficiency of conversion of energy 
 in these objects should significantly exceed the "nominal" 10 percent value.

\end{methods}

\vfill
\clearpage

\renewcommand{\refname}{References}

 \renewcommand{\figurename}{Supplementary Figure}
\setcounter{figure}{0}

\begin{figure*}
\centering
\includegraphics[width=0.48\linewidth]{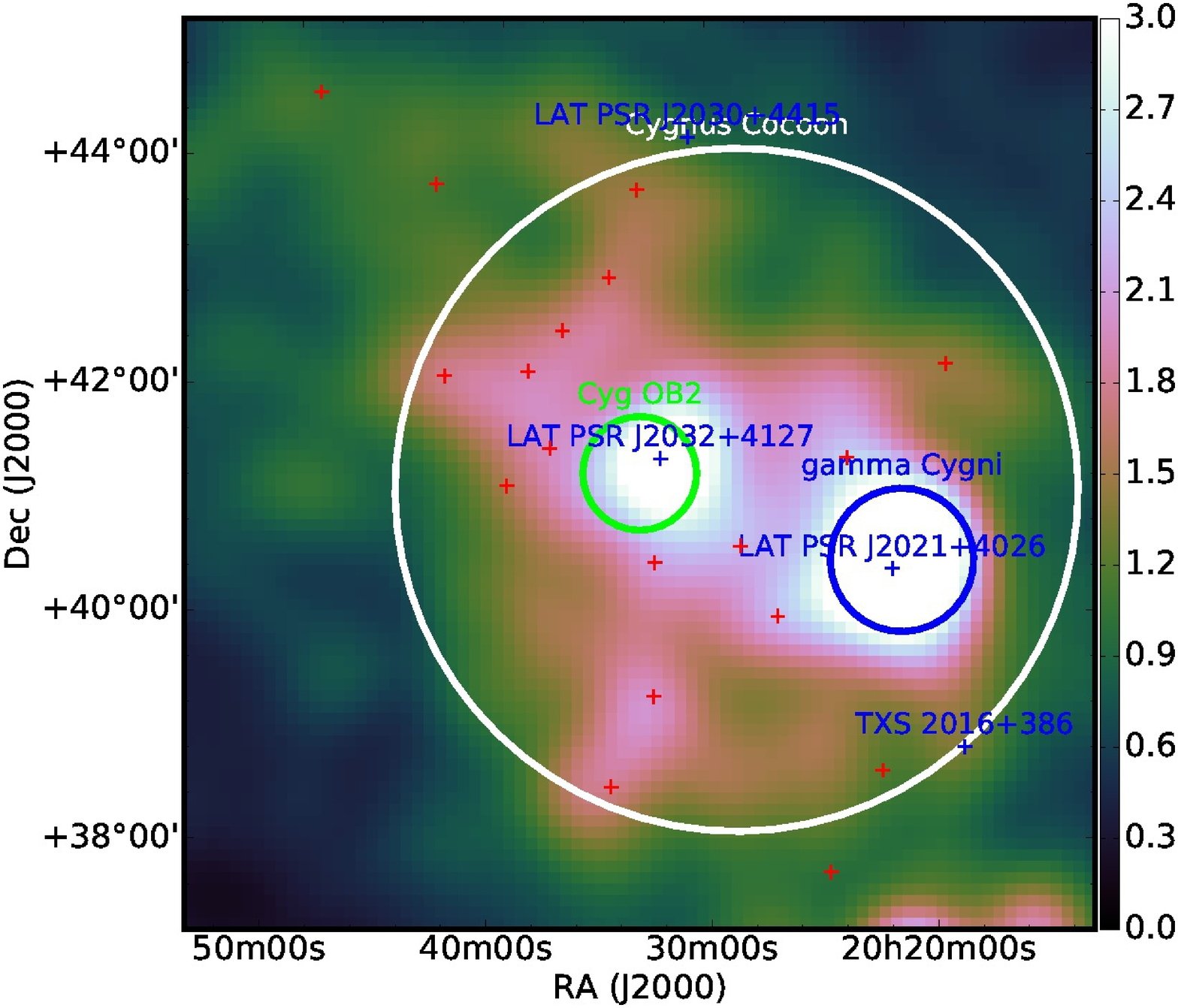}
\includegraphics[width=0.48\linewidth]{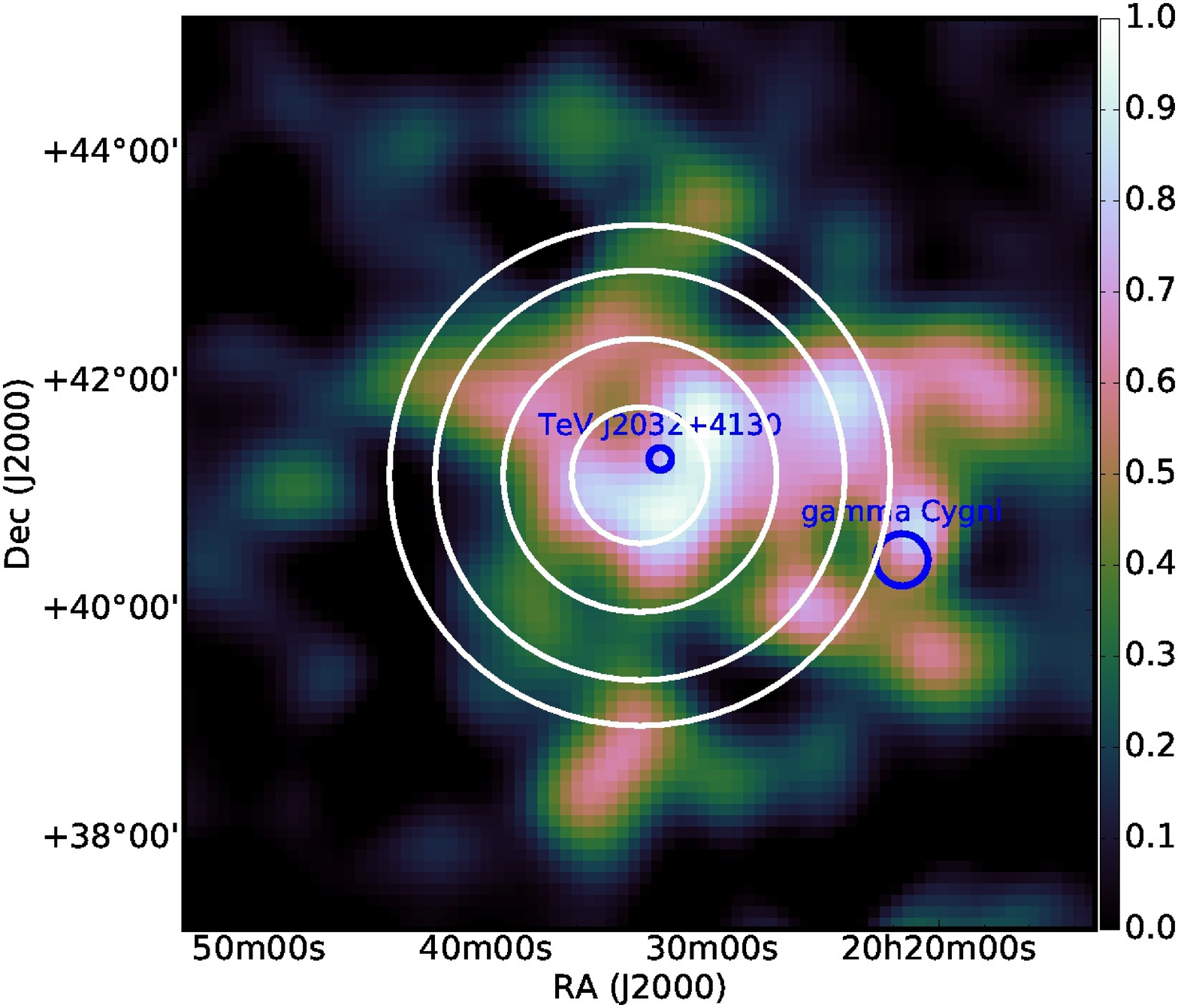}
\caption{The counts (left panel) and residual (right panel) maps of $\gamma$-rays around the Cygnus Cocoon at the 
 energies above 10~GeV.  The color bars are in counts per pixel. A  $10^\circ \times 10^\circ$ region around Cygnus Cocoon is shown.  
The positions of the 3FGL catalog sources are marked   with 
blue and red crosses for the identified and unassociated 
{\it Fermi} LAT sources, respectively.  Also shown is the position of the extended source Gamma Cygni (blue circle) and the position of young star association Cygnus OB2 (green circle).  The residual map is obtained 
after subtracting all identified catalog sources as well as the 
diffuse backgrounds.  The white circles 
on the residual map represent  the regions  used  for the extraction of 
the radial distribution of $\gamma$-ray emissivities. The blue circles show the position and extension in the TeV range of two TeV sources Gamma Cygni and TeV J2032+415. }
\label{fig:cmap}
\end{figure*}

 \begin{figure*}
\centering
\includegraphics[width=0.4\linewidth]{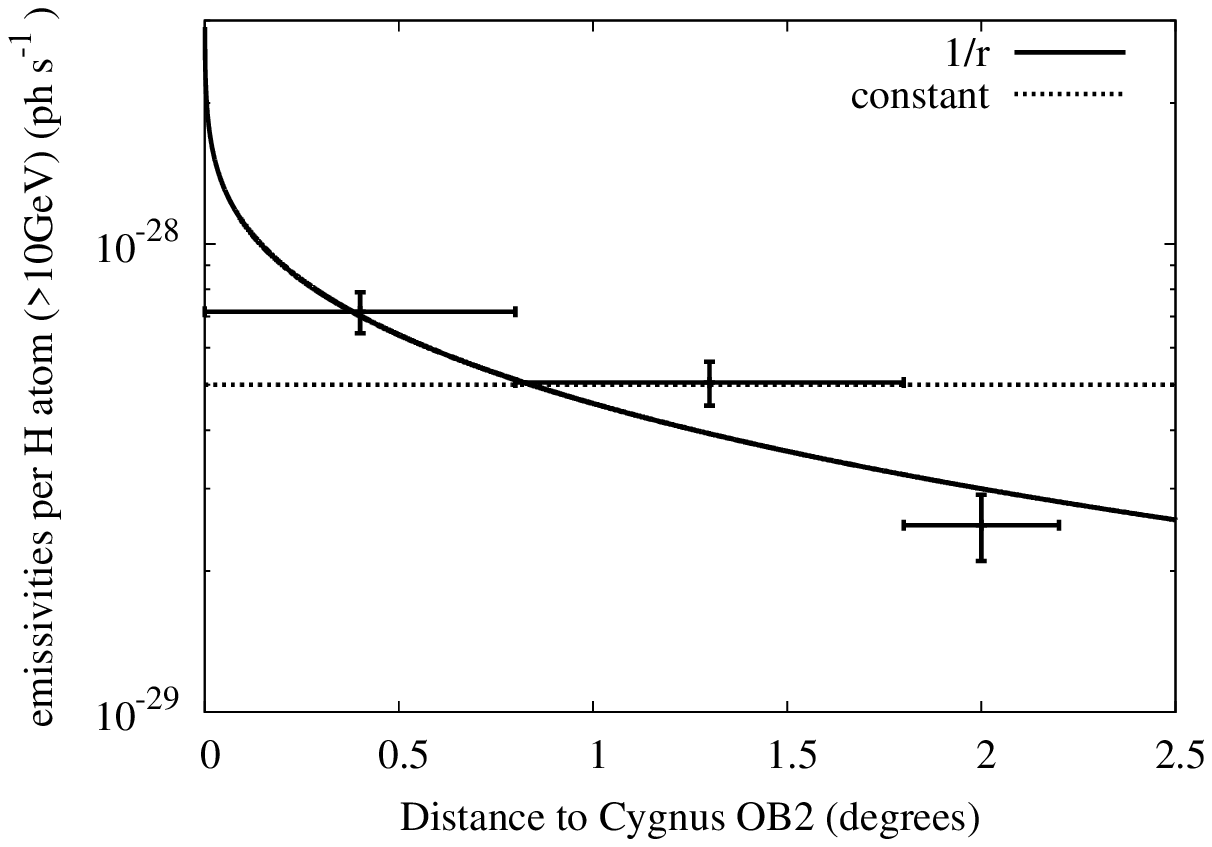}
\includegraphics[width=0.4\linewidth]{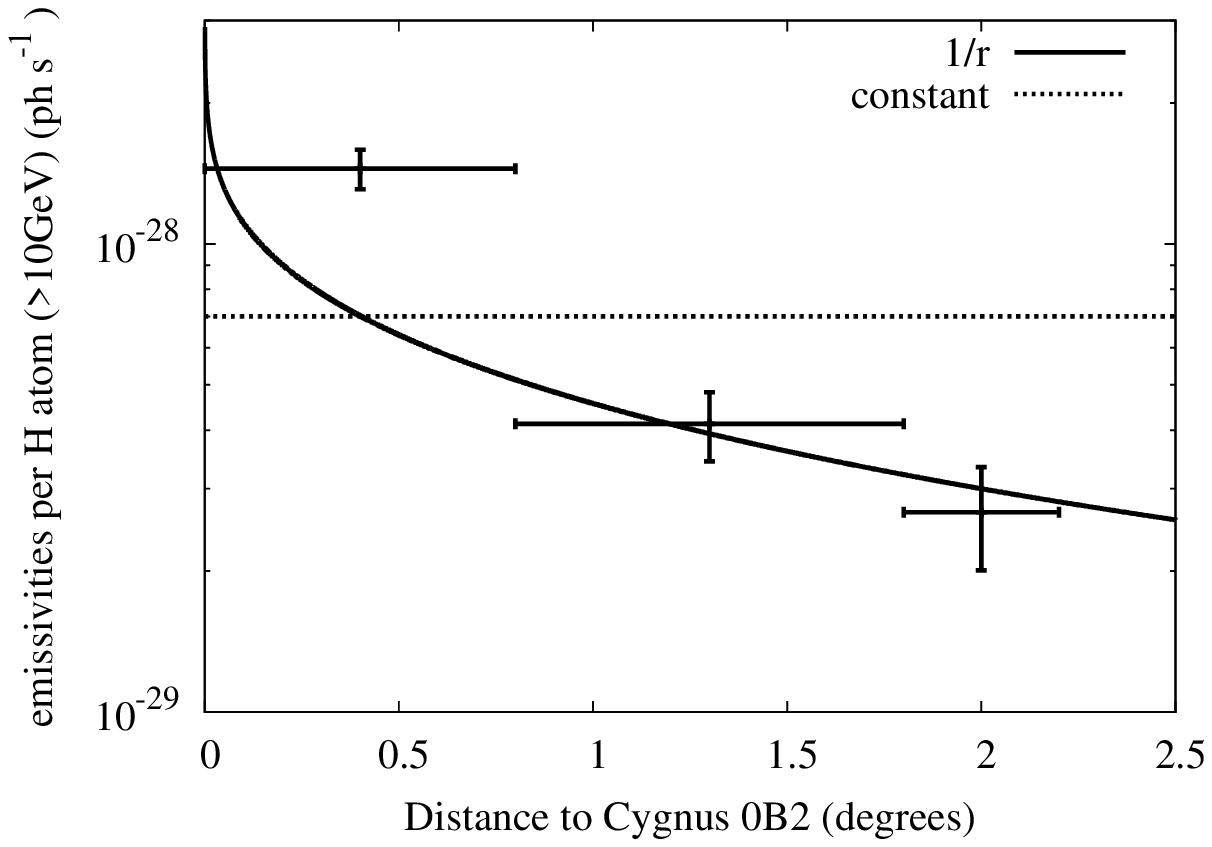}
\caption{ The radial profiles of  $\gamma$-ray emissivities (per H-atom)  above 10~GeV with respect to the position of  Cygnus OB2 in the south (left panel) and north (right panel) hemisphere of Cygnus Cocoon.  The error bars contain both statistical and systematic errors.}
\label{fig:pro_div}
\end{figure*}

%
%
%

\begin{figure*}
\centering
\includegraphics[width=0.34\linewidth]{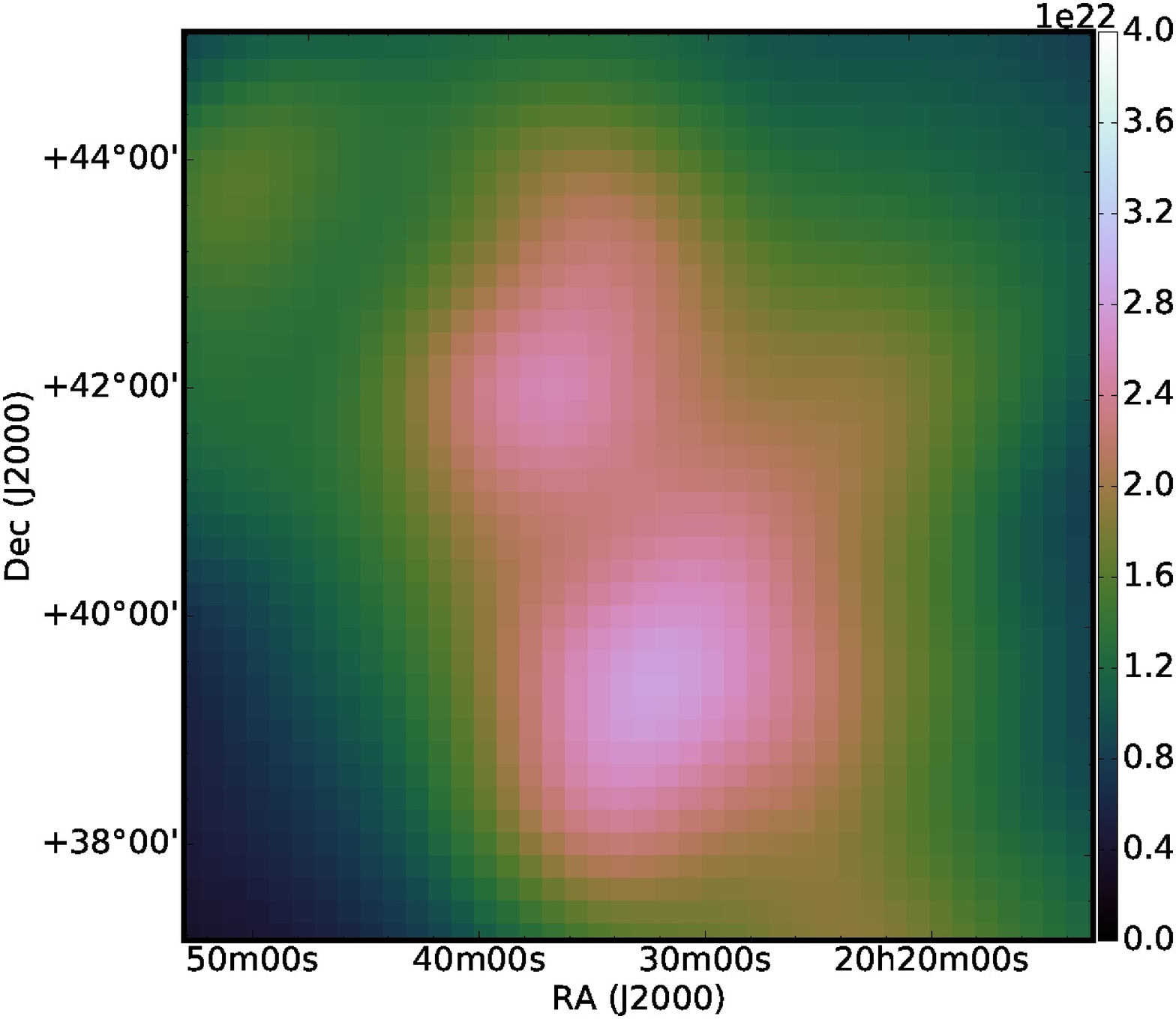}
\includegraphics[width=0.34\linewidth]{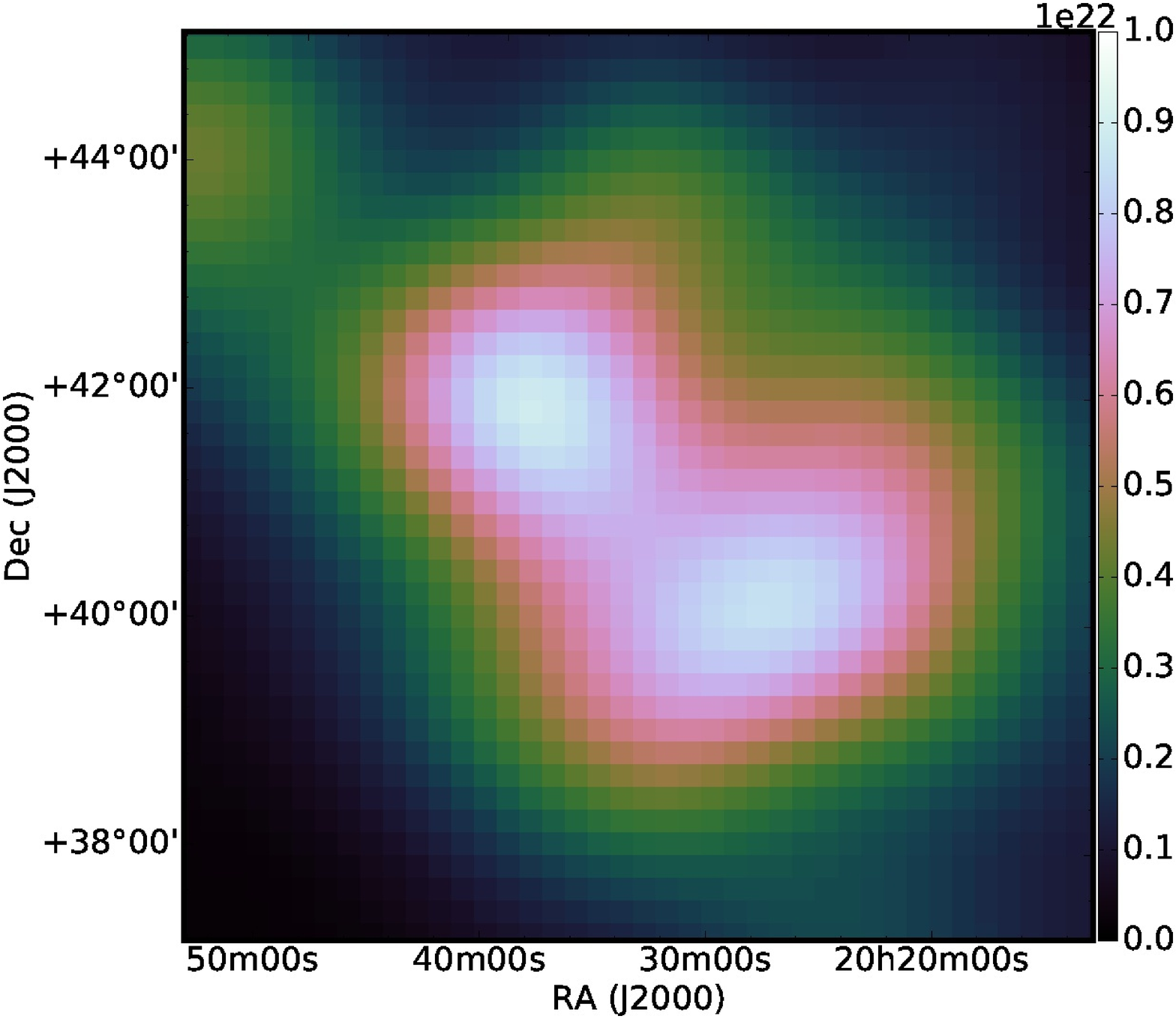}
\includegraphics[width=0.34\linewidth]{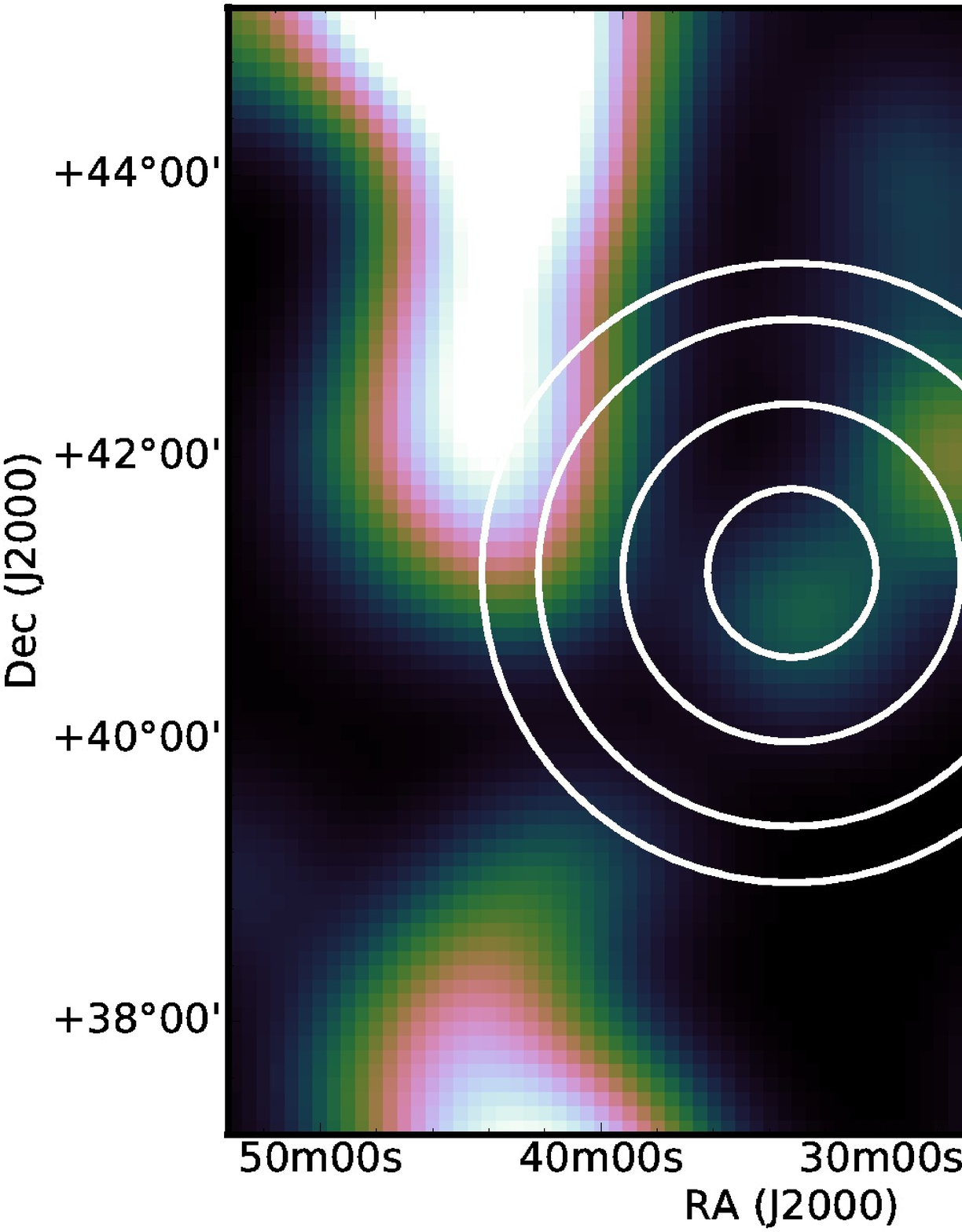}

\caption{{\it left top}: the molecular and neutral atomic hydrogen column density  in the Local Arm towards Cygnus Cocoon.  {\it right top}. The ionised hydrogen density derived assuming for the electron density $n_e = 2 ~\rm cm^{-3}$. {\it bottom}. Dark gas distribution in Cygnus region derived using the method described in Sec.4 in Method.  The white circles  represent  the regions  used  for the extraction of 
the radial distribution of $\gamma$-ray emissivities.  The color bars show gas column density in the unit of $\rm cm^{-2}$.  }.
\label{fig:gas}
\end{figure*}

%

\begin{figure*}
\centering
\includegraphics[width=0.6\linewidth]{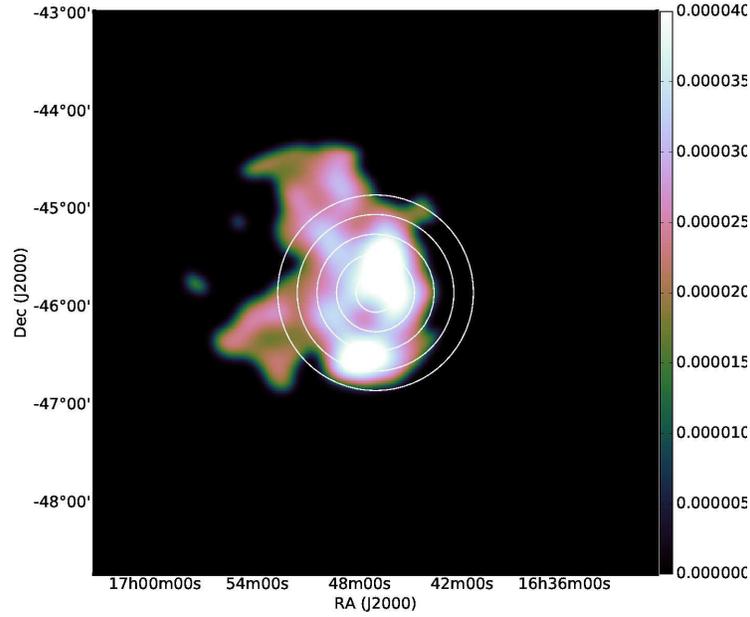}

\caption{ The HESS VHE excess map for Westerlund 1 region. The map was produced using the public data in H.E.S.S website. The white circles  represent  the regions  used  for the extraction of 
the radial distribution of $\gamma$-ray emissivities. The color bars are in units of  excess per pixel. }
 \label{fig:hess_w1}
\end{figure*}

\begin{figure*}
\centering
\includegraphics[width=0.4\linewidth]{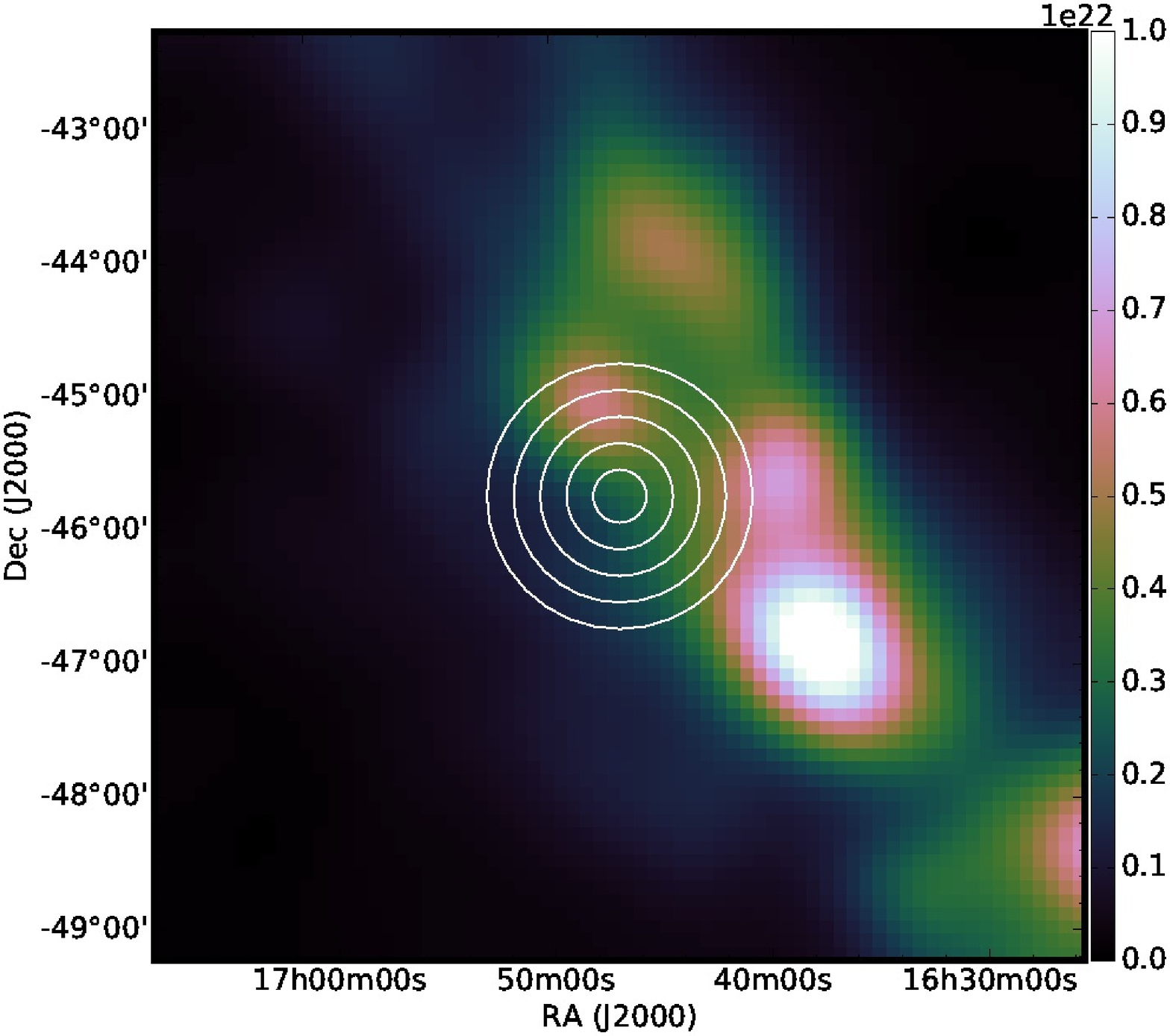}
\includegraphics[width=0.4\linewidth]{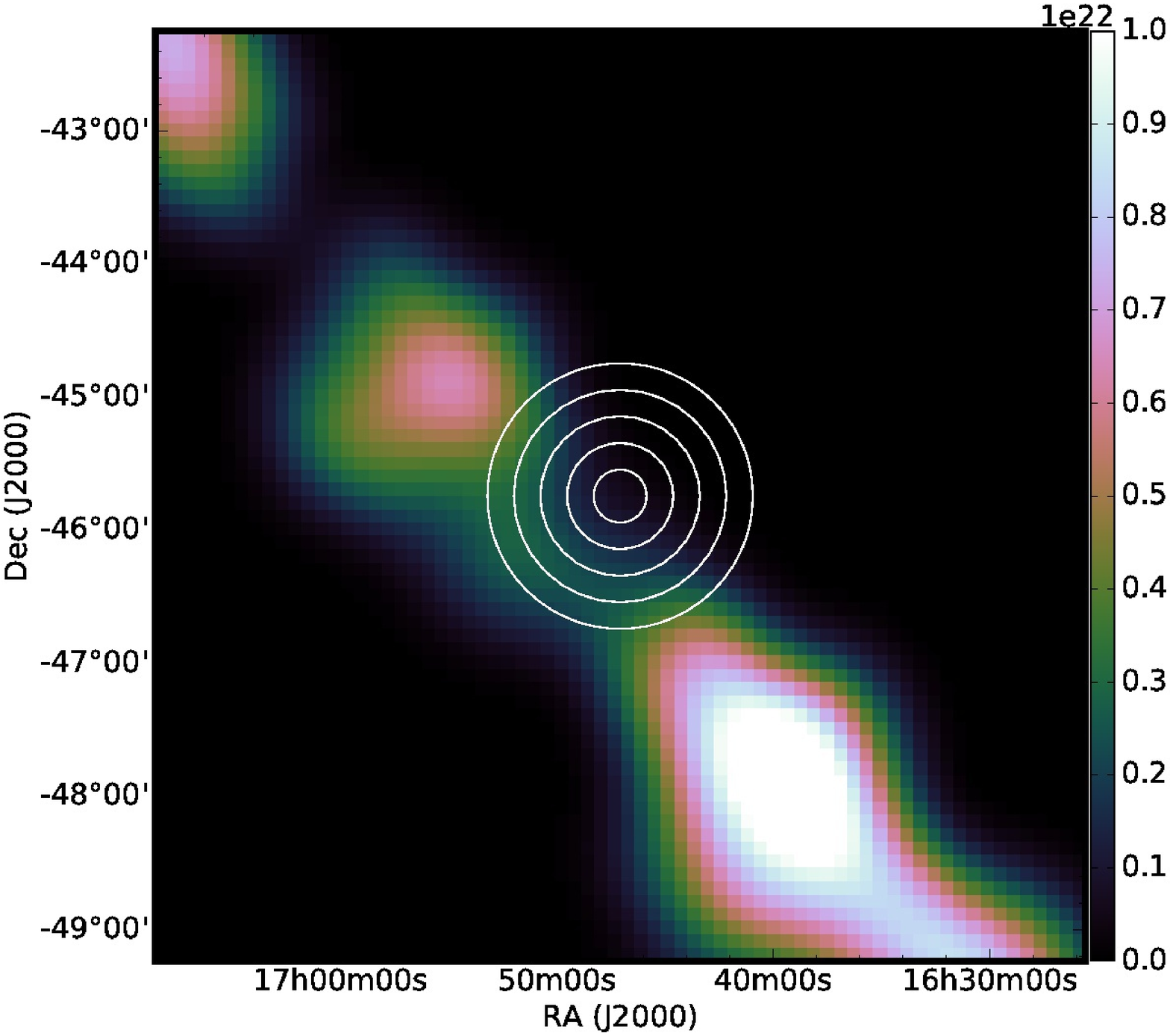}

\caption{ {\it Left panel}: The molecular and neutral atomic hydrogen column density in the vicinity of Westerlund 1. The velocity range  $-60  ~\rm km/s <V_{\rm LSR} <-50 ~\rm km/s$ are integrated to derive the gas distribution. The green rings label the region to subtract the radial distribution of CRs. {\it right panel}: Dark gas distribution near Westerlund 1 derived using the method described in Sec.4 in Method.    The white circles  represent  the regions  used  for the extraction of 
the radial distribution of $\gamma$-ray emissivities.  The color bars show gas column density in the unit of $\rm cm^{-2}$. }
 \label{fig:gas_w1}
\end{figure*}
\begin{figure}
\centering
\includegraphics[width=1\linewidth]{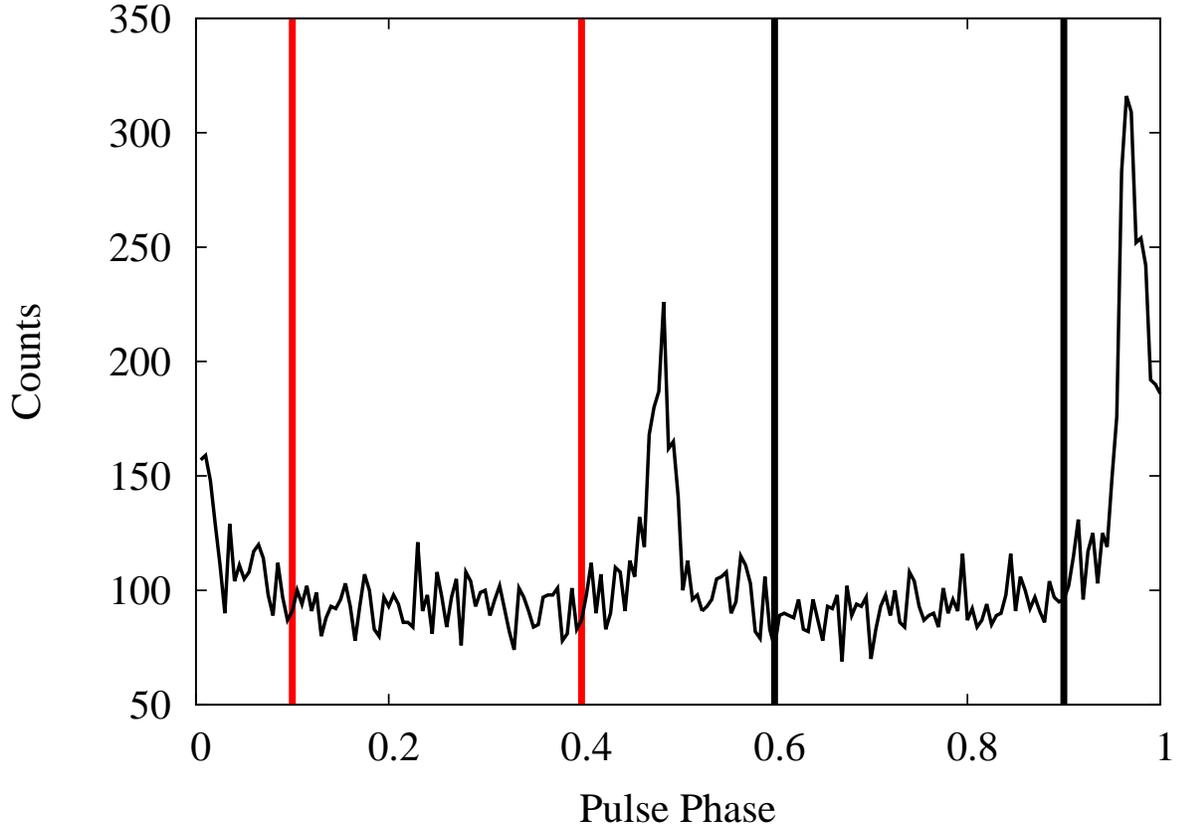}
\caption{Phase-folded light curve of the pulsar LAT PSR J2032+4127. The $\gamma$-ray data 
have been selected, for the further analysis,  only for the intervals between two black and two 
read vertical lines.}
\label{fig:lc}
\end{figure}

 \begin{figure*}
\centering
\includegraphics[width=0.4\linewidth]{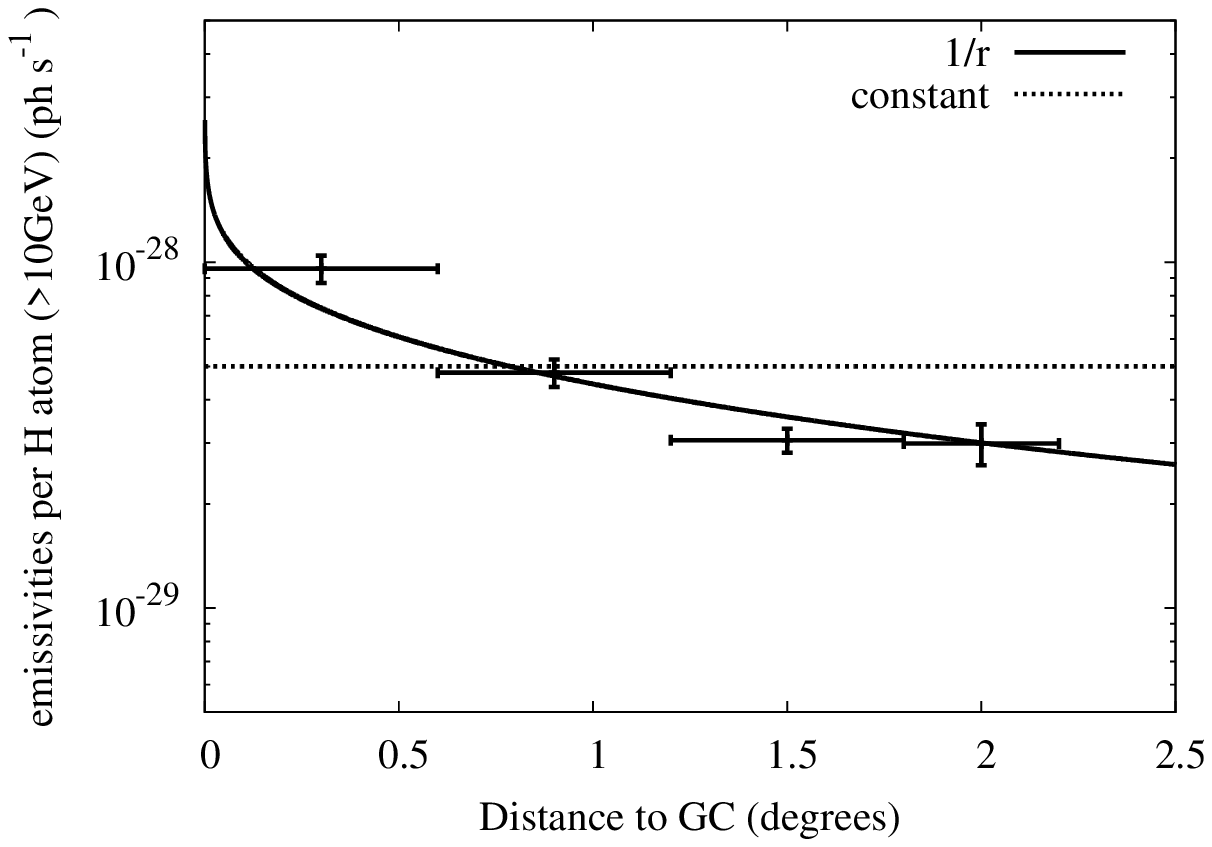}
\includegraphics[width=0.4\linewidth]{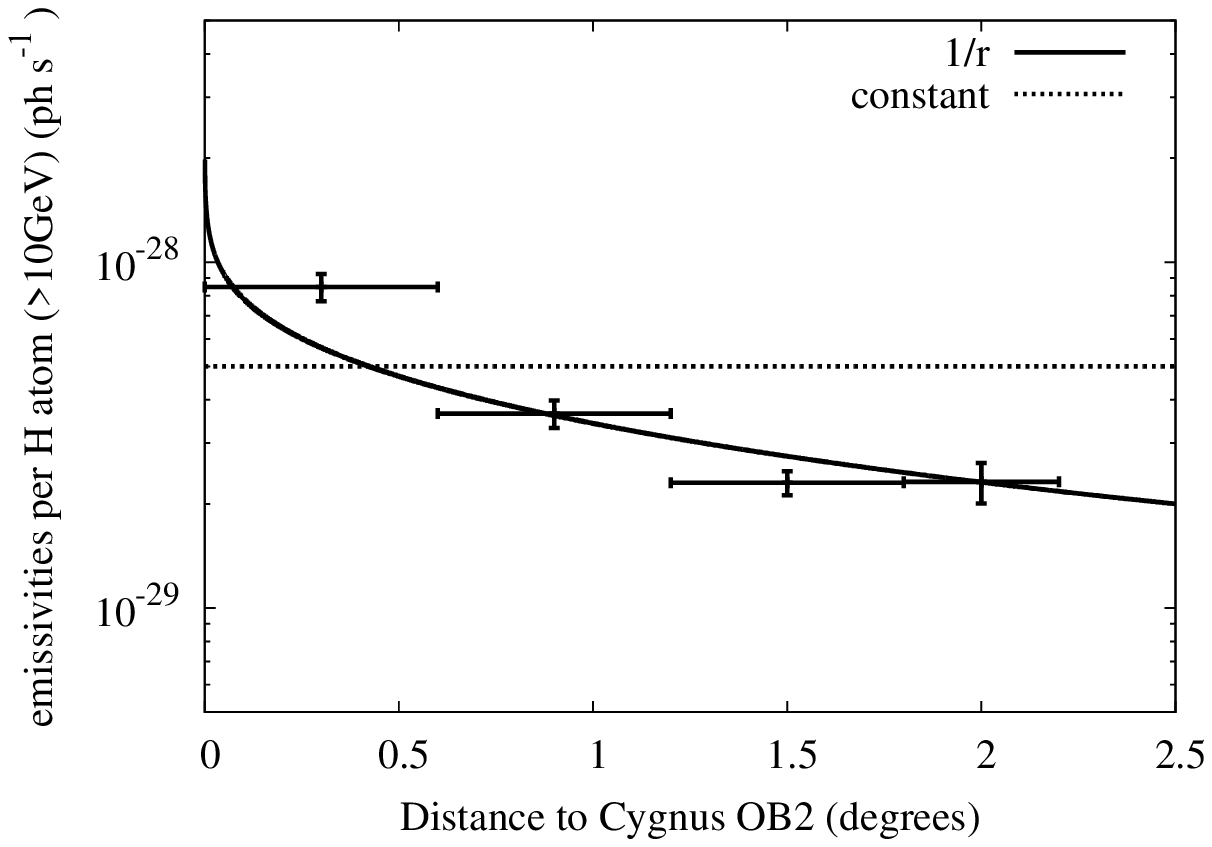}

\caption{ Left panel: The profile of $\gamma$-ray emissivities (per H-atom) above 10~GeV with respect to the location of   Cygnus OB2. For comparison, we show $1/r$ (solid curve), and constant (dotted curve) profiles, which are expected in the cases of  continuous  and impulsive injections, respectively.   
The radial distributions are shown for  all three  (atomic, molecular and ionised) components of gas 
involved in the $\gamma$-ray production. Right panel: The same as the left panel but the "dark gas" component are also included to derive the CR distribution.  The error bars contain both statistical and systematic errors. }
\label{fig:pro_cygnus}
\end{figure*}

 \begin{figure*}
\centering
\includegraphics[width=0.4\linewidth]{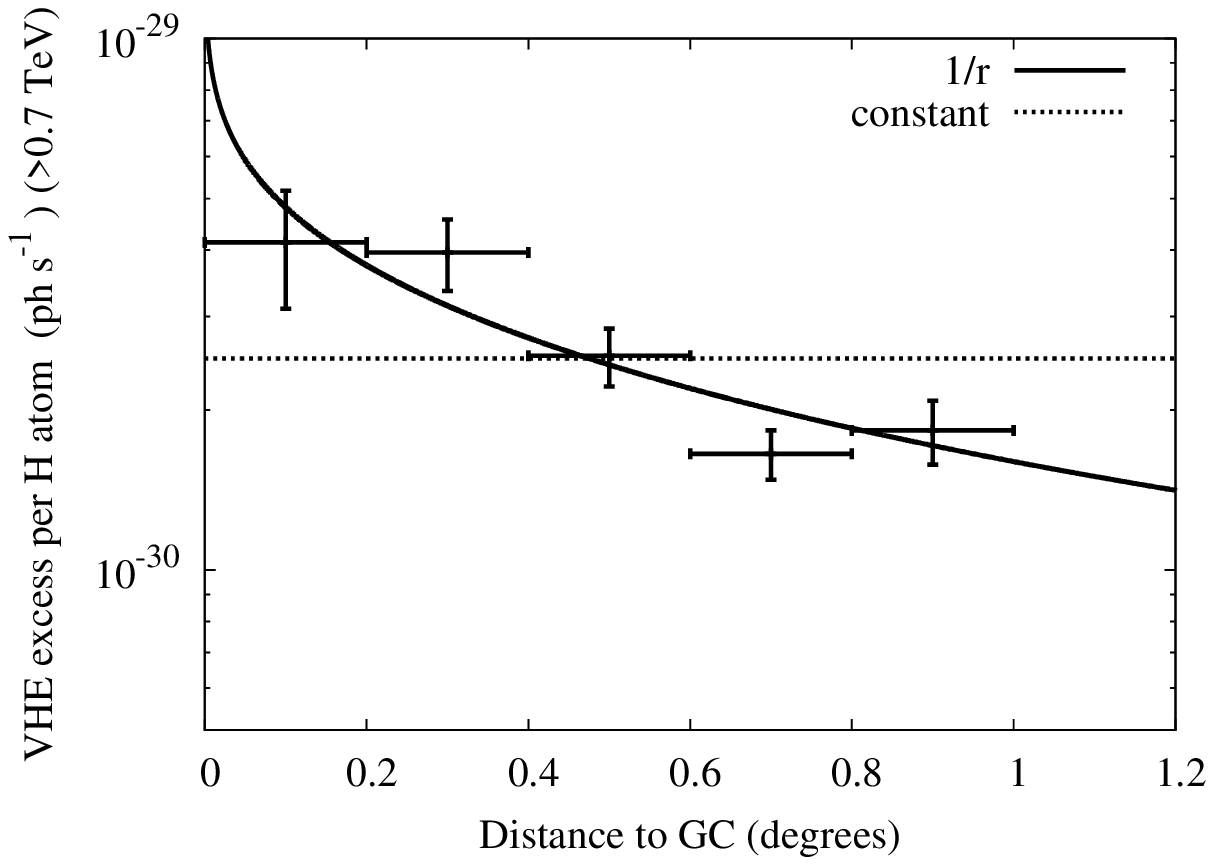}
\includegraphics[width=0.4\linewidth]{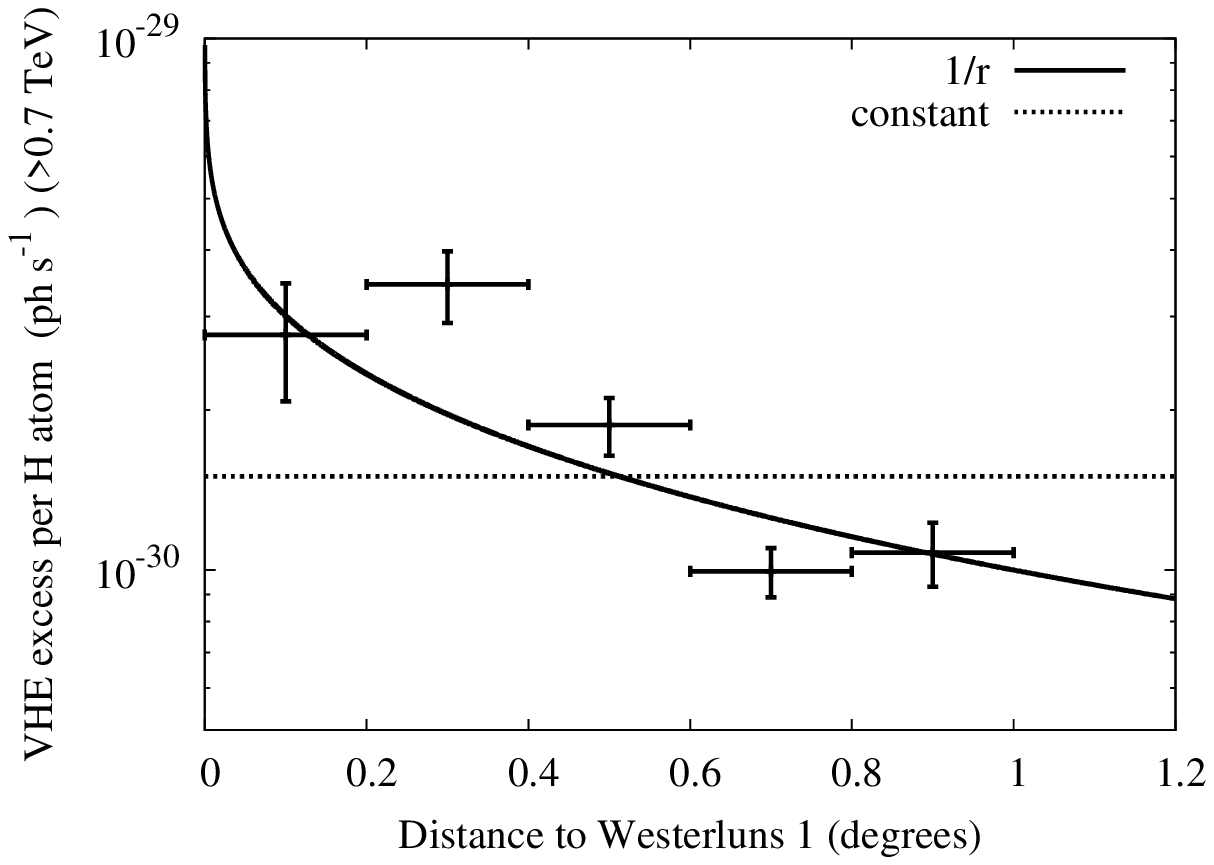}

\caption{ Left panel: The VHE excess per  H-atom in the vicinity of Westerlund 1. The curves are the same as those in Supplementary Figure\ref{fig:pro_cygnus}. Right panel: The same as the left panel but the "dark gas" component are also included to derive the CR distribution. The error bars contain both statistical and systematic errors. }
\label{fig:pro_w1}
\end{figure*}

 \begin{figure*}
\centering
\includegraphics[width=0.4\linewidth]{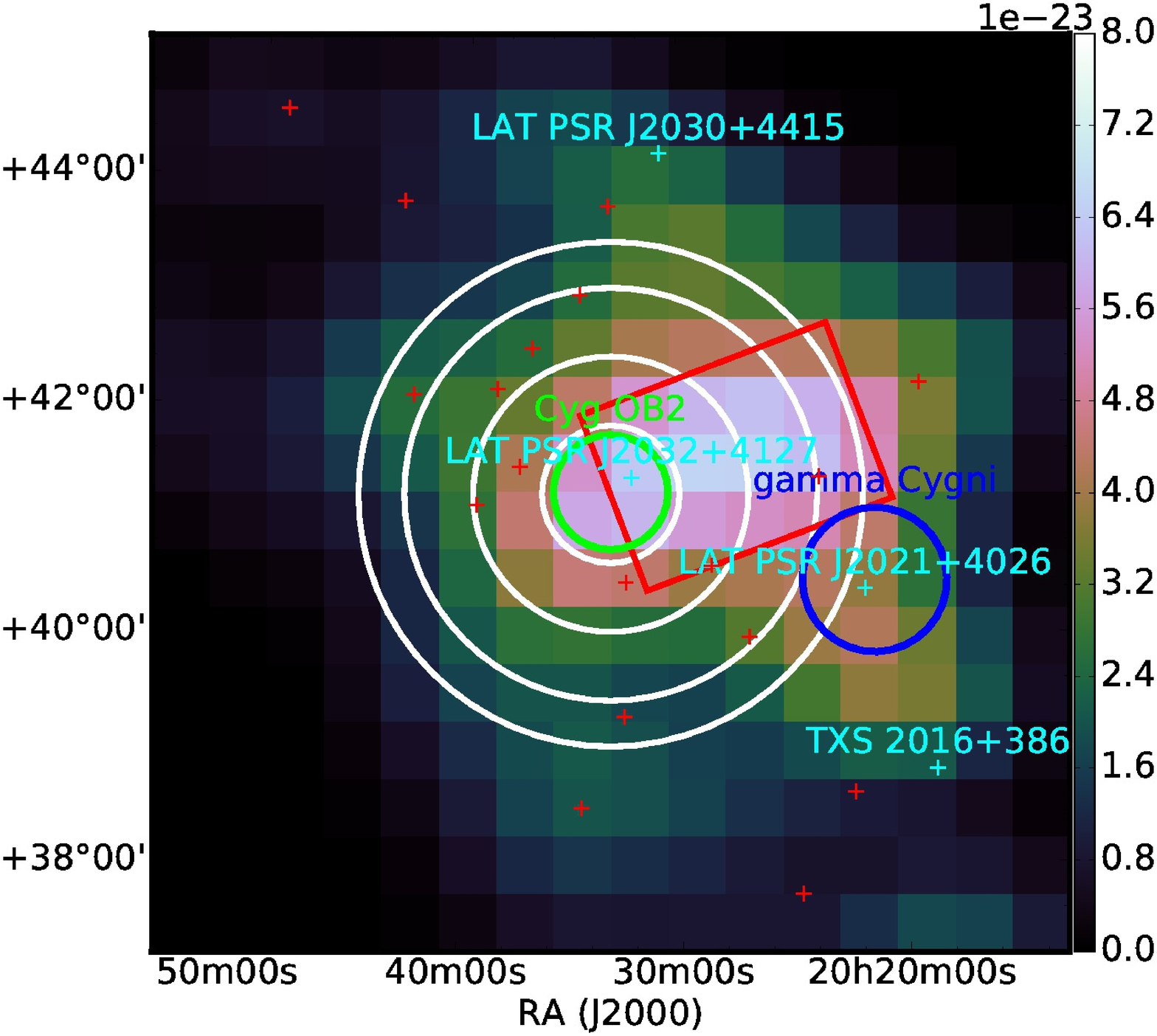}
\includegraphics[width=0.4\linewidth]{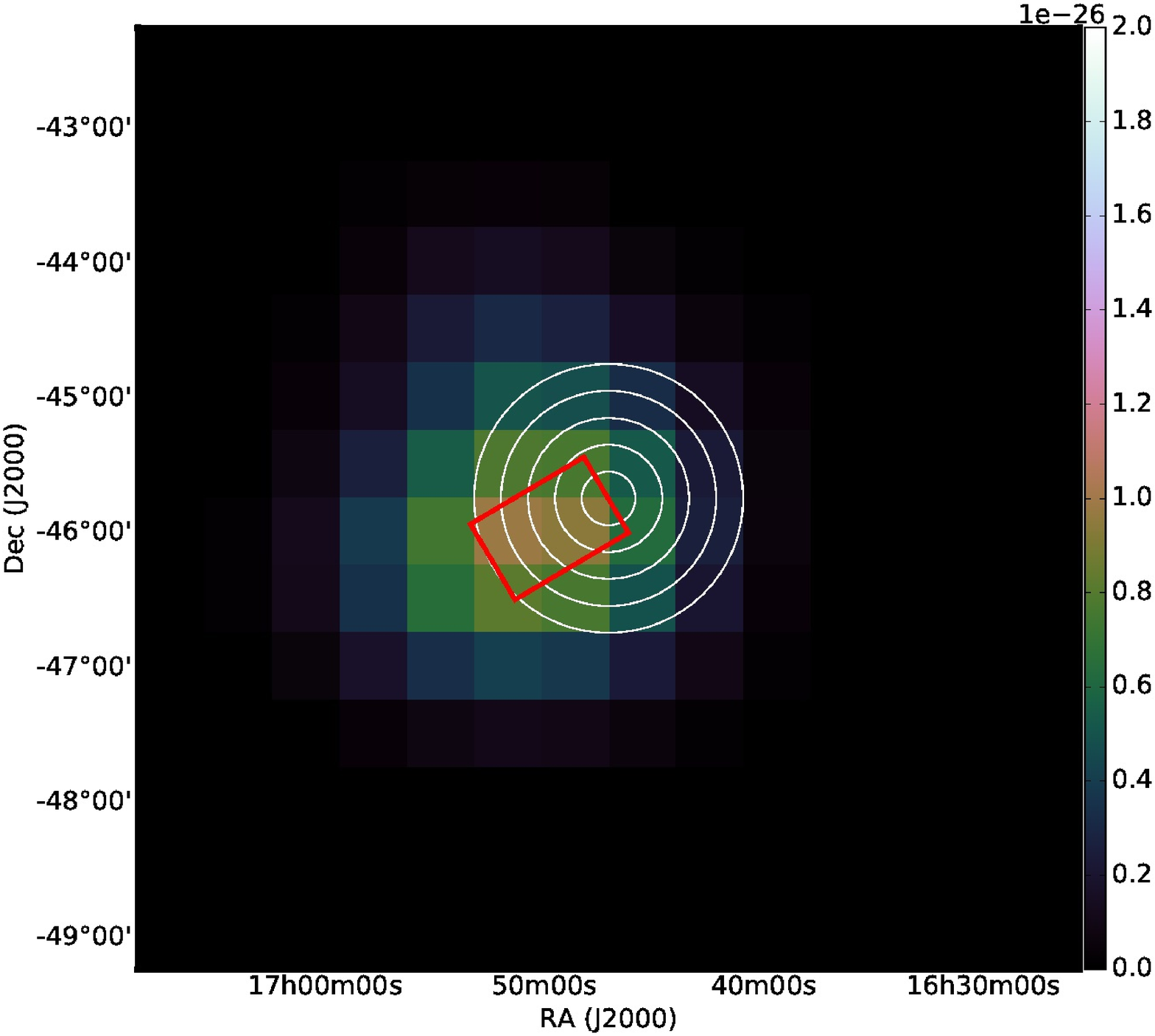}\\
\includegraphics[width=0.4\linewidth]{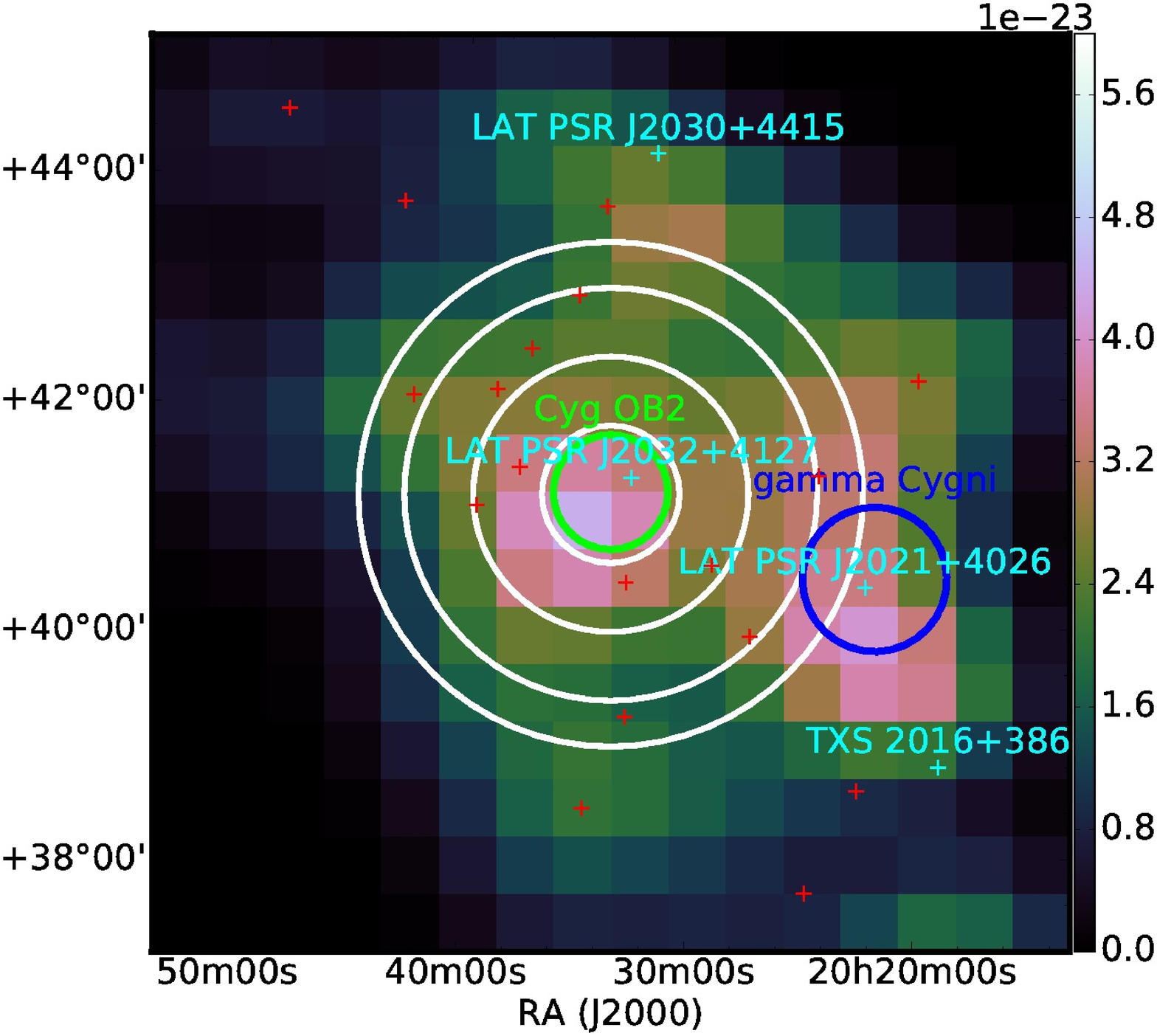}

\caption{ Top: The normalized emissivity maps for Cygnus Cocoon (top left panel) and Westerlund 1 (top right panel). The normalized emissivity map is derived by dividing the Wd 1 HESS excess map and Cygnus Cocoon Fermi LAT residual map by the corresponding gas maps, respectively. The areas indicated by red boxes are the regions with significant excess caused, most likely, by background or foreground extended sources. Bottom: The normalized emissivity maps for Cygnus Cocoon after removing the emission corresponding to the excess templates in the likelihood fitting.   The white circles  represent  the regions  used  for the extraction of 
the radial distribution of $\gamma$-ray emissivities.  The color bars are  in the unit of  $\rm counts/excess per pixel \times cm^2$.  }
\label{fig:emis}
\end{figure*}

 \begin{figure*}
\centering
\includegraphics[width=0.4\linewidth]{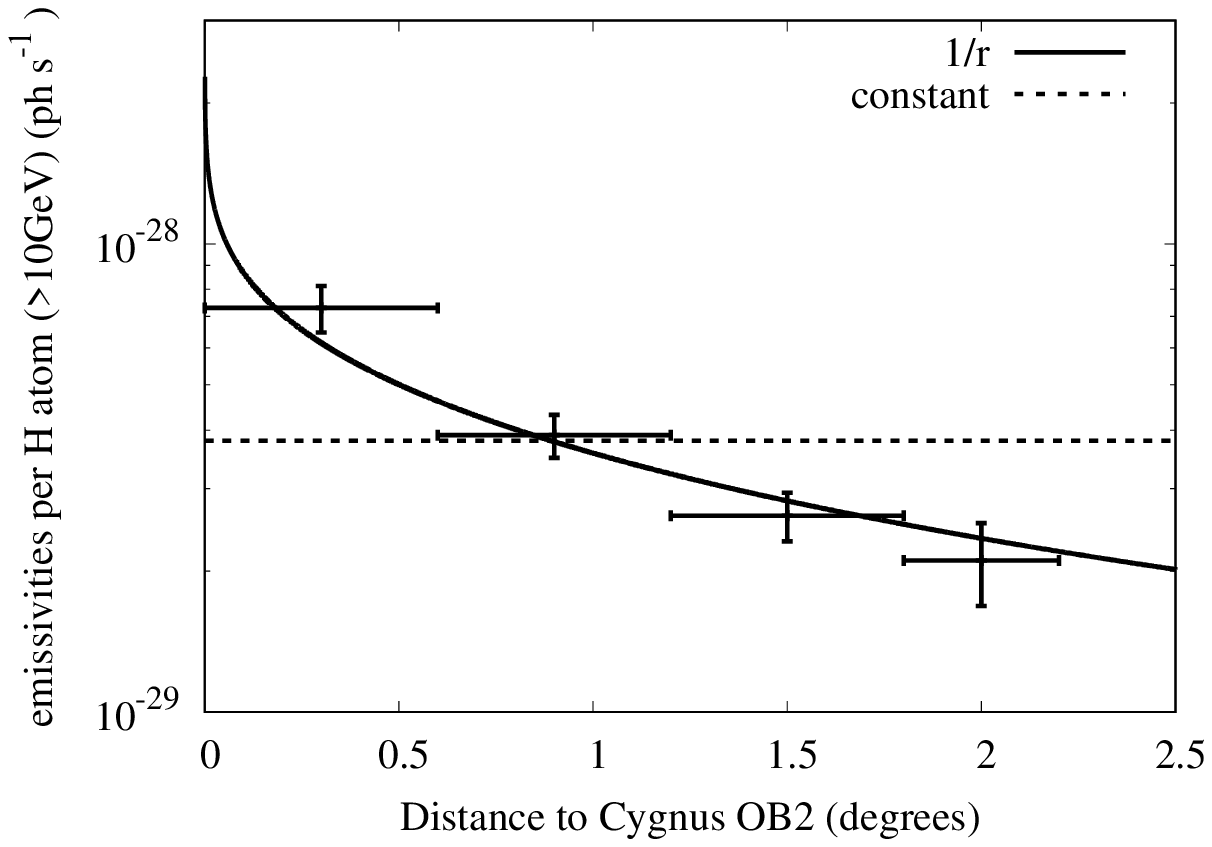}
\includegraphics[width=0.4\linewidth]{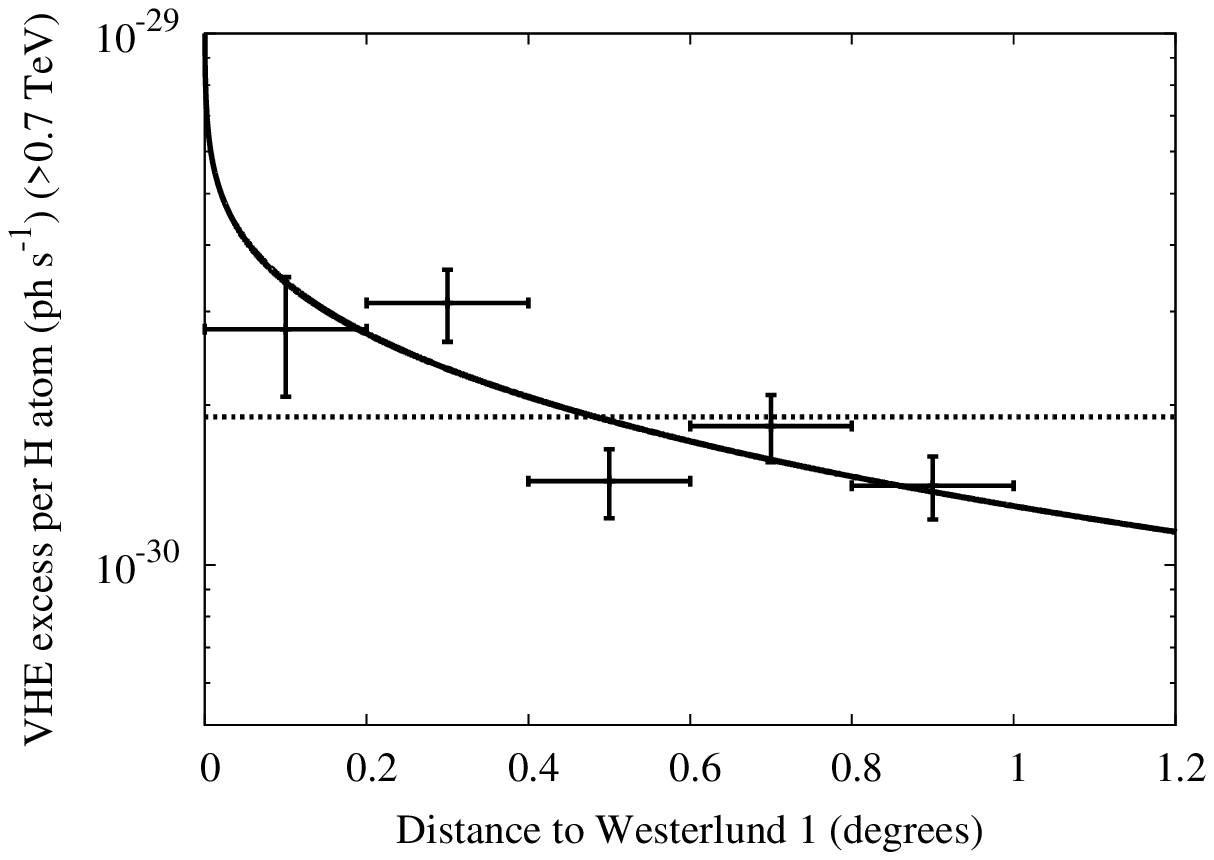}

\caption{  The profile of $\gamma$-ray emissivities (per H-atom) above 10~GeV for the regions around Cygnus OB2 (left panel) and Westerlund 1 (right panel) after subtraction of the external excess emission contained in the red boxes in Supplementary Figure \ref{fig:emis}. The error bars contain both statistical and systematic errors.}
\label{fig:noex}
\end{figure*}

\begin{figure*}
\centering
\includegraphics[width=0.4\linewidth]{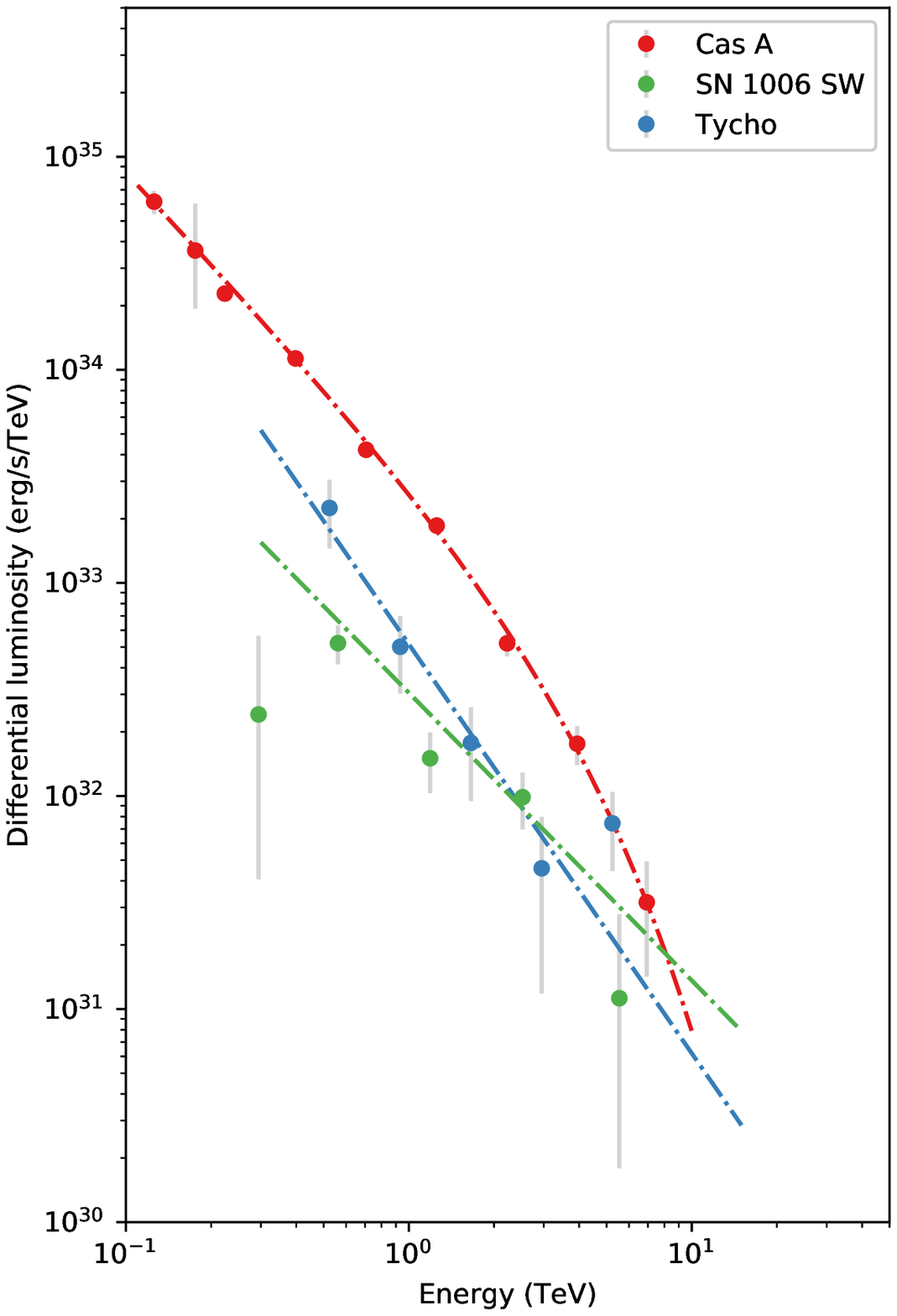}
\includegraphics[width=0.4\linewidth]{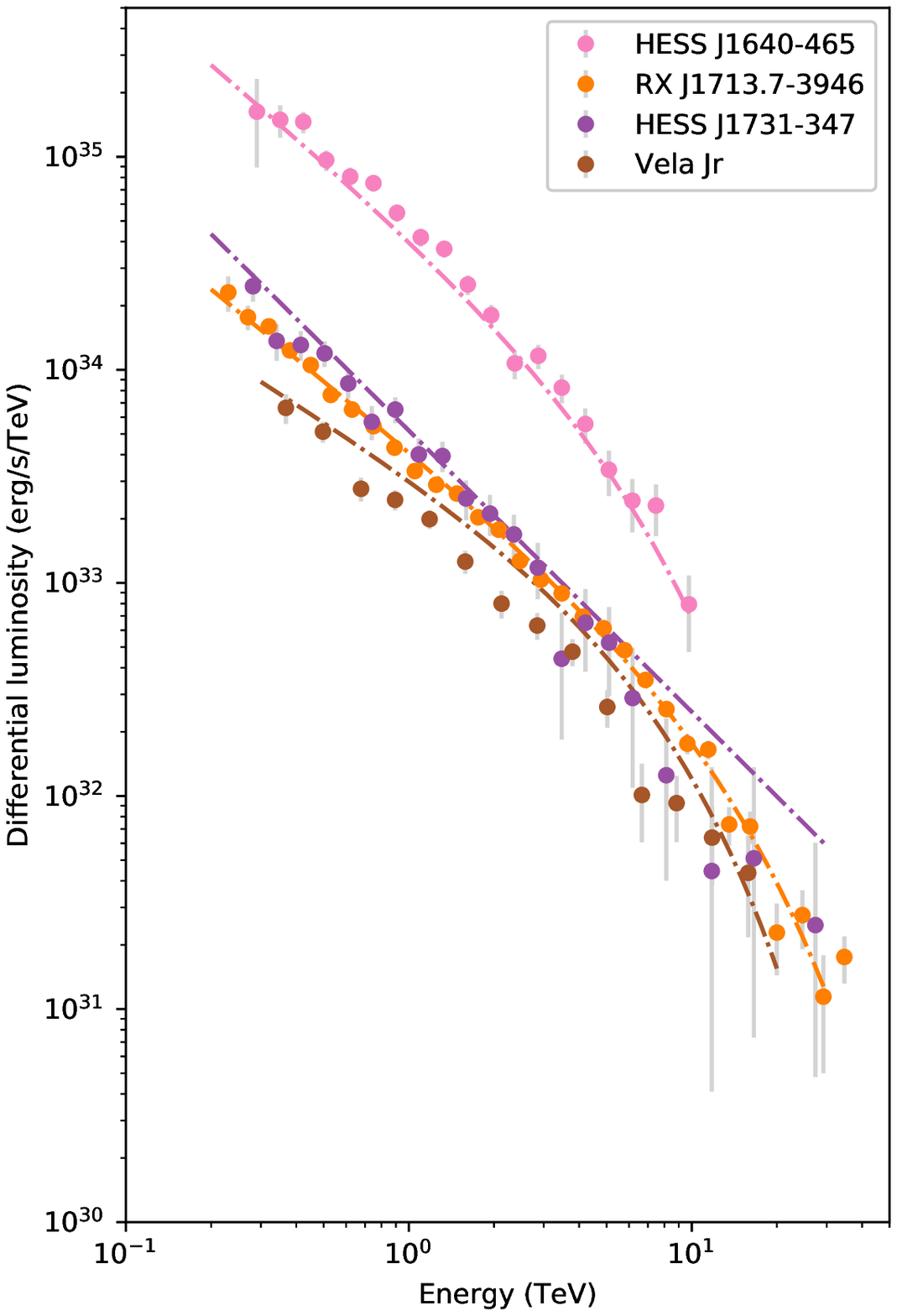}
\caption{Differential spectrum of historical (left panel) and other (right panel) TeV emitting SNRs. }
\label{fig:TeVSNRs}
\end{figure*}

\clearpage

\renewcommand{\tablename}{Supplementary Table}

\newpage
\setcounter{table}{0}   
\begin{table}
\begin{tabular}{|c|c|c|c|c|c|}
\hline
 ~Region ~ & ~~$L_{\gamma}$ ($>$10 GeV) ($10^{34} \rm erg/s$)~~& mass ($10^5~\rm M_{\rm \odot}$)  & $w_{\rm CR}$  ($>100 \rm ~GeV) (eV/cm^3)$ & index & significance ($\sigma$) \\
 \hline
$0 ~\rm pc<r <15~\rm pc$ & $0.60 \pm 0.06$  & 0.8 &$ 0.13\pm 0.013$&$2.3\pm0.2$ & 12.2\\
$15~\rm pc<r <29~\rm pc$  & $0.87 \pm 0.09$ & 2.4 & $0.065 \pm 0.007$&$2.0\pm0.2$ & 11.2\\
$29 ~\rm pc<r <44~\rm pc$ & $0.91 \pm 0.13$	& 4.0 &$0.041 \pm 0.006$&$2.0\pm0.2$ & 10.1\\
$44 ~\rm pc<r <54~\rm pc$  & $ 0.64\pm 0.07$	   & 3.3 & $0.035 \pm 0.004$&$2.3\pm0.2$ & 8.3\\
\hline
\end{tabular}
\caption{$\gamma$-ray luminosities, gas masses, CR densities, $\gamma$-ray spectral indices and detection significance in different regions of the Cygnus Cocoon.  The error bars contain both statistical and systematic errors.}
\end{table}


\begin{table}
\begin{tabular}{|c|c|c|c|c|}
\hline
 ~~~~Region ~ & ~~$L_{\gamma} $($>$1 TeV) ($10^{34} \rm erg/s$)~~& mass ($10^5~\rm M_{\rm \odot}$)  & $w_{\rm CR}  (>10 \rm ~TeV) (eV/cm^3)$\\
 \hline
$ 0~\rm pc<r <13~\rm pc$ & $0.41\pm 0.10$  & 0.1 &$ 0.73\pm 0.18$\\
$13~\rm pc<r <26~\rm pc$  & $1.18 \pm 0.27$ & 0.3 & $0.71 \pm 0.16$ \\
$26 ~\rm pc<r <39~\rm pc$ & $1.87 \pm 0.20$	& 0.73&$0.46 \pm 0.05$\\
$39~\rm pc<r <52~\rm pc$  & $ 2.01\pm 0.27$	   & 1.20& $0.30 \pm 0.04$\\
$52~\rm pc<r <65~\rm pc$  & $ 1.92\pm 0.23$	   & 1.03& $0.33\pm 0.04$\\

\hline
\end{tabular}
\caption{$\gamma$-ray luminosities, gas masses and CR densities in different regions of the Wd1 Cocoon.  The error bars contain both statistical and systematic errors.}
\end{table}

\clearpage

%
%


\end{document}